\documentclass[letterpaper]{raa}           
\usepackage{graphicx,times}
\usepackage{natbib}
\usepackage{amssymb,amsmath}
\usepackage{soul}
\bibpunct{(}{)}{;}{a}{}{,}
\usepackage[pagebackref=true]{hyperref}

\newcommand{\bs}[1]{\boldsymbol{#1}}
\newcommand{\code}[1]{\texttt{#1}}
\newcommand{\dd}{\mathrm{d}}
\newcommand{\NN}{\mathrm{N}}

\begin{document}

\title{Modelling the covariance matrix for the power spectra before and after the BAO reconstruction}

\volnopage{ {\bf 20XX} Vol.\ {\bf X} No. {\bf XX}, 000--000}
\setcounter{page}{1}

\author{
Ruiyang Zhao\inst{1,2,3}
\and
Kazuya Koyama\inst{3}
\and
Yuting Wang\inst{1,4}
\and
Gong-Bo Zhao\inst{1,2,4,5}
}

\institute{
National Astronomical Observatories, Chinese Academy of Sciences, Beijing, 100101, P.R.China\\
\and
School of Astronomy and Space Science, University of Chinese Academy of Sciences, Beijing 100049, P.R.China\\
\and
Institute of Cosmology \& Gravitation, University of Portsmouth, Dennis Sciama Building, Portsmouth, PO1 3FX, UK\\
\and
Institute for Frontiers in Astronomy and Astrophysics, Beijing Normal University, Beijing, 102206, P.R.China\\
\and
Chinese Academy of Sciences South America Center for Astronomy (CASSACA), National Astronomical Observatories of China, Beijing, 100101, P.R.China\\
\vs \no
{\small Received 20XX Month Day; accepted 20XX Month Day}
}

\abstract{The baryon acoustic oscillation (BAO) reconstruction plays a crucial role in cosmological analysis for spectroscopic galaxy surveys because it can make the density field effectively more linear and more Gaussian. The combination of the power spectra before and after the BAO reconstruction helps break degeneracies among parameters, then improve the constraints on cosmological parameters. It is therefore important to estimate the covariance matrix between pre- and post-reconstructed power spectra. In this work, we use perturbation theory to estimate the covariance matrix of the related power spectra multipoles, and check the accuracy of the derived covariance model using a large suite of dark matter halo catalogs at $z=0.5$. We find that the diagonal part of the auto covariance is well described by the Gaussian prediction, while the cross covariance deviates from the Gaussian prediction quickly when $k > 0.1\,h\,\mathrm{Mpc}^{-1}$. Additionally, we find the non-Gaussian effect in the non-diagonal part of the cross covariance is comparable to, or even stronger than the pre-reconstruction covariance. By adding the non-Gaussian contribution, we obtain good agreement between analytical and numerical covariance matrices in the non-diagonal part up to $k \simeq 0.15\,h\,\mathrm{Mpc}^{-1}$. The agreement in the diagonal part is also improved, but still under-predicts the correlation in the cross covariance block.
}

\authorrunning{R. Zhao et al. }            %author_head in even pages
\titlerunning{Analytic covariance}  % title_head in odd pages
\maketitle

\section{Introduction}
\label{sec:intro}
The large-scale structure of the Universe carries rich information of the cosmic expansion, the structure growth and primordial physics, and the information leaves imprints on patterns of the three-dimensional distribution of galaxies such as the Baryon Acoustic Oscillations \citep[BAO;][]{SDSS:2005xqv,2dFGRS:2005yhx} and the Redshift Space Distortions \citep[RSD;][]{Kaiser:1987qv, Peacock:2001gs}, as measured by current and future spectroscopic surveys including Dark Energy Spectroscopic Instrument \citep[DESI;][]{DESI:2016fyo}, Euclid \citep{Euclid:2011}, Subaru Prime Focus Spectrograph \citep[PFS;][]{PFS:2014}, Nancy Grace Roman Space Telescope \citep[Roman;][]{Roman:2021oec}, MegaMapper \citep{MegaMapper:2022vrv}, and so forth. Given the large number of galaxies, the observational data is usually compressed and analysed at the 2-point statistics level such as the $2$-point correlation function and the power spectrum. The $2$-point statistics capture the leading order information of the galaxy density field, and have been well modelled in theory \citep[e.g., ][]{Crocce:2005xy, Taruya:2010mx, Baumann:2010tm, Carrasco:2012cv, Vlah:2015sea, Maus:2024sbb}, and measured and analysed over the last decades \citep[e.g., ][]{Beutler:2011hx, BOSS:2016wmc, DAmico:2019fhj, Ivanov:2019pdj,eBOSS:2020yzd,DESI:2024uvr,DESI:2024mwx}.

To fully exploit the information from galaxy surveys, there have been attempts to measure and model high-order statistics \citep{Gil-Marin:2016wya,Pearson:2017wtw,Sugiyama:2018yzo,Desjacques:2018pfv,Eggemeier:2018qae,Gualdi:2021yvq,Philcox:2021hbm,Philcox:2022frc,DAmico:2022osl,Novell-Masot:2023rli,Spaar:2023his,Wang:2023zkv,SimBIG:2023nol,Sugiyama:2023zvd,Behera:2023uat,Leonard:2024xvy,Chen:2024pyp}. These analyses are, however, much more challenging compared to the 2-point statistics because of the high dimensionality of the observable.

The reconstruction technique \citep{Eisenstein:2006nj} was proposed to improve the BAO measurement by undoing the large-scale bulk flow which blurred the BAO feature. Although it was originally designed to improve the BAO signal, reconstruction, which is in effect a field level operation, utilizes the higher-order information \citep{Schmittfull:2015mja} to restore the linear modes that are contaminated by the nonlinear gravitational evolution. Therefore, cosmological information, such as RSD \citep{Zhu:2017vtj, Hikage:2019ihj}, primordial non-Gaussianity \citep{Shirasaki:2020vkk, Floss:2023ylq} and the neutrino mass \citep{Zang:2023rpx}, can in principle be better constrained by analysing the post-reconstruction sample compared to the pre-reconstruction one when stopped at the same order of $n$-point statistics. Over the past few years, theoretical studies of the reconstructed field have been mainly focusing on modelling the signal \citep[see][]{White:2015eaa, Seo:2015eyw, Hikage:2017tmm, Hikage:2019ihj, Chen:2019lpf, Ota:2021caz, Chen:2024tfp, Sugiyama:2024eye, Sugiyama:2024ggt, Sugiyama:2024qsw}. 
Recently, \cite{Wang:2022nlx, Wang:2023hlx} show the combination of pre and post reconstructed BAO or power spectra breaks degeneracies among parameters and significantly improves the constraints on cosmological parameters. In this case, it is also important to understand the covariance matrix between pre and post reconstructed power spectra because it is nearly singular at small $k$.

The standard way to estimate the covariance matrix is based on numerical simulations. There are, however, a few disadvantages. First, the finite number of simulations introduces noise into the numerical covariance matrix. The matrix is likely to be ill-formed when observables are highly correlated. Moreover, the noise in the inverse covariance matrix estimation has to be account for \citep{Hartlap:2006kj} and be propagated to model parameters \citep{Percival:2013sga, Percival:2021cuq}, weakening the final cosmological constraints. Second, limited to the computational resources, running a large number of full N-body simulations is usually not available and one has to rely on fast approximated mocks \citep[e.g., ][]{Kitaura:2015uqa, Feng:2016yqz, eBOSS:2020wwo}. Although they are usually well calibrated at the 2-pt statistics level, it is not entirely clear how much they are able to match the higher order statistics \citep{Philcox:2024mmz}. On the other hand, the perturbation theory provides a clean way to understand all related effects in the covariance matrix calculation \citep{Scoccimarro:1999kp, Meiksin:1998mu, Bertolini:2015fya, Mohammed:2016sre, Chan:2016ehg, Sugiyama:2019ike, Wadekar:2019rdu}. It also provides the ability to perform the quick cross-check with the numerical covariance matrix \citep{Wadekar:2020hax}.

This paper extends the results in \citet{Hikage:2020fte} to include the galaxy bias and the discreteness effect based on perturbation theory and compute the analytic covariance matrix. In addition, we also model the cross covariance matrix between pre- and post-reconstruction power spectra. We compare the analytic covariance against numerical ones computed from $15000$ Quijote simulations \citep{Villaescusa-Navarro:2019bje}.

This paper is organised as follows. Section \ref{sec:model} presents the analytic covariance matrix model. In Section \ref{sec:dem}, we compare our analytic covariance matrix model to the numerical one computed from $15000$ Quijote simulations. The conclusions and discussions are summarized in Section \ref{sec:con}.

\section{The covariance matrix model}
\label{sec:model}
\subsection{Density fluctuations and correlators}
\label{sec:density-fluctuations}
Following the counts in the cell formalism in \citet{Feldman:1993ky}, \citet{Smith:2008ut} and \citet{Sugiyama:2024ggt}, we can divide the simulation box into infinitesimal cells with volume $\delta V$. The number of galaxies\footnote{Our derivations in this section also apply to haloes, so we will use ``galaxy'' and ``halo'' interchangeably in the later section.} in the $i$-th cell, or the galaxy occupation number $n_{g,i}$, is either $0$ or $1$, satisfying
\begin{equation}
n_{g,i} = n_{g,i}^n 
\label{eq:ng_i}
\end{equation}
where $n \geq 2$. The galaxy number density in the simulation box is then expressed as
\begin{equation}
n_g(\bs{x}) = \sum_i n_{g,i} \delta_D(\bs{x} - \bs{x}_i)\text{,}
\end{equation}
with $\delta_D$ the Dirac delta function and $\bs{x}_i$ the position of the $i$-th cell. The expectation value of the occupation number is $\langle n_{g,i} \rangle = \bar{n}_g \delta V$, with $\bar{n}_g$ the galaxy number density. Assuming galaxies are distributed following the Poisson point process, the occupation numbers in different cells ($i\not=j\not=k\not=l$) are only correlated through the underlying continuous galaxy density field
\begin{align}
\langle n_{g, i} n_{g, j} \rangle_c &= \bar{n}_g^2 \xi_{gg}(\bs{x}_i, \bs{x}_j) \delta V^2 \\
\langle n_{g, i} n_{g, j} n_{g, k} \rangle_c &= \bar{n}_g^3 \zeta_{ggg}(\bs{x}_i, \bs{x}_j, \bs{x}_k) \delta V^3\\
\langle n_{g, i} n_{g, j} n_{g, k} n_{g, l} \rangle_c &= \bar{n}_g^4 \eta_{gggg}(\bs{x}_i, \bs{x}_j, \bs{x}_k, \bs{x}_l) \delta V^4\text{,}
\end{align}
where we use notation $\langle \cdots \rangle_c$ to represent the cumulant average, and we have defined the two-, three- and four-point galaxy correlation functions $\xi_{gg}, \zeta_{ggg}$ and $\eta_{gggg}$, respectively.

In Fourier space, the galaxy number density becomes
\begin{equation}
n_{g}(\bs{k}) = \sum_i n_{g,i} e^{-i \bs{k} \cdot \bs{x}_i}\text{.}
\end{equation}
For the convenience of later calculations, we can define
\begin{equation}
\delta_{g,i}(\bs{k}) = \frac{1}{\bar{n}_g} n_{g,i} e^{-i \bs{k} \cdot \bs{x}_i}\text{.}
\end{equation}
The summation of $\delta_{g,i}$ satisfies
\begin{equation}
\sum_i \langle \delta_{g,i}(\bs{k}) \rangle_c = \frac{1}{\bar{n}_g} \sum_i \langle n_{g,i} \rangle_c e^{-i \bs{k} \cdot \bs{x}_i} = \sum_i e^{-i \bs{k} \cdot \bs{x}_i} \delta V
\rightarrow \int_V e^{-i \bs{k} \cdot \bs{x}} \dd^3 x = V \delta^K_{\bs{k}}\text{,}
\end{equation}
where $V$ is the volume of simulation box, and $\delta^{K}_{\bs{k}}$ is the Kronecker delta symbol, which is equal to $1$ when $\bs{k}=\bs{0}$ and vanishes otherwise. Here the arrow indicates that we take the continuous limit.
The $2$-point correlation of $\delta_{g,i}$ is
\begin{align}
&\sum_{i \not= j} \langle \delta_{g,i}(\bs{k}_1) \delta_{g,j}(\bs{k}_2) \rangle_c = \sum_{i\not= j} \xi_{gg}(\bs{x}_i, \bs{x}_j) e^{-i \bs{k}_1 \cdot \bs{x}_i} e^{-i \bs{k}_2 \cdot \bs{x}_j}\delta V^2\\
\to &\int_V e^{-i (\bs{k}_1 + \bs{k}_2) \cdot \bs{x}_2} \dd^3 x_2 \int_V \xi_{gg}(|\bs{x}_1 - \bs{x}_2|) e^{-i \bs{k}_1 \cdot (\bs{x}_1 - \bs{x}_2)} \dd^3|\bs{x}_1 - \bs{x}_2| = V\delta^K_{\bs{k}_{12}} P_{gg}(\bs{k}_1)\text{,}
\end{align}
where $P_{gg}$ is the power spectrum, and we use the notation $\bs{k}_{12} \equiv \bs{k}_1 + \bs{k}_2$. Similarly, higher order correlations of $\delta_{g,i}$ are related to bispectrum and trispectrum, respectively,
\begin{align}
\sum_{i\not=j\not=k}  \langle \delta_{g,i}(\bs{k}_1) \delta_{g,j}(\bs{k}_2) \delta_{g,j}(\bs{k}_3) \rangle_c &= V\delta^K_{\bs{k}_{123}} B_{ggg}(\bs{k}_1, \bs{k}_2, \bs{k}_3)\\
\sum_{i\not=j\not=k\not=l}  \langle \delta_{g,i}(\bs{k}_1) \delta_{g,j}(\bs{k}_2) \delta_{g,j}(\bs{k}_3) \delta_{g,l}(\bs{k}_4) \rangle_c &= V\delta^K_{\bs{k}_{1234}} T_{gggg}(\bs{k}_1, \bs{k}_2, \bs{k}_3, \bs{k}_4)\text{.}
\end{align}

\subsection{The shot noise effect}
Correlation of occupation numbers in the same cell is called the discreteness effect or the shot noise effect. It can be evaluated using eq(\ref{eq:ng_i}), or equivalently
\begin{equation}
\delta_{g,i} (\bs{k}_1) \delta_{g,i} (\bs{k}_2) = \frac{1}{\bar{n}_{g}} \delta_{g,i} (\bs{k}_{12}) \text{.} \label{eq:delta_gi_contract}
\end{equation}
We will use the superscript $\NN$ to represent correlators with the shot noise effect, i.e.,
\begin{align}
\sum_{i,j}  \langle \delta_{g,i}(\bs{k}_1) \delta_{g,j}(\bs{k}_2) \rangle_c &= V\delta^K_{\bs{k}_{12}} P^\NN_{gg}(\bs{k}_1)\\
\sum_{i,j,k}  \langle \delta_{g,i}(\bs{k}_1) \delta_{g,j}(\bs{k}_2) \delta_{g,k}(\bs{k}_3) \rangle_c &= V\delta^K_{\bs{k}_{123}} B^\NN_{ggg}(\bs{k}_1, \bs{k}_2, \bs{k}_3)\\
\sum_{i,j,k,l}  \langle \delta_{g,i}(\bs{k}_1) \delta_{g,j}(\bs{k}_2) \delta_{g,k}(\bs{k}_3) \delta_{g,l}(\bs{k}_4) \rangle_c &= V\delta^K_{\bs{k}_{1234}} T^\NN_{gggg}(\bs{k}_1, \bs{k}_2, \bs{k}_3, \bs{k}_4)\text{.}
\end{align}
The following expression is also useful in the later calculation
\begin{equation}
\sum_{i\not=j, k}  \langle \delta_{g,i}(\bs{k}_1) \delta_{g,j}(\bs{k}_2) \delta_{g,k}(\bs{k}_3) \rangle_c = V\delta^K_{\bs{k}_{123}} B^{\NN12}_{ggg}(\bs{k}_1, \bs{k}_2, \bs{k}_3)\text{.}
\end{equation}
Here the superscript $\mathrm{N}12$ indicates that the correlation between the first and the second density field in the same cell is removed, i.e., the shot noise effect is partially removed in this correlator. Their explicit expressions are given in Appendix~\ref{appendix:perturbation}.

\subsection{The BAO Reconstruction}
In the standard Zeldovich reconstruction, the density field after reconstruction $n_{*}$ reads
\begin{equation}
n_{*}(\bs{x}) = n_d(\bs{x}) - \alpha n_s(\bs{x})\text{,}
\end{equation}
where $\alpha$ is the ratio of the total number of galaxies to randoms, $n_d$ is the displaced data number density, and $n_s$ is the displaced random number density. They are obtained by moving particles in the galaxy catalogue $n_g$ and an initially unclustered catalogue consist of random particles $n_r$ according to the shift field $\bs{s}(\bs{x})$. The shift field in Fourier space can be explicitly written as
\begin{equation}
\bs{s}(\bs{k})=i\bs{R}(\bs{k})\delta_{g}(\bs{k})\text{.} \label{eq:shift-field}
\end{equation}
Here $\delta_g$ is the galaxy overdensity field $\delta_g(\bs{x}) = n_g(\bs{x}) / \bar{n}_g - 1$, while kernel $\bs{R}$ determines the reconstruction detail
\begin{equation}
\bs{R}(\bs{k})=-\frac{\bs{k}+f_{\mathrm{fid}}(\bs{k}\cdot\hat{\bs{\eta}})\hat{\bs{\eta}}}{k^{2}}\frac{\mathcal{S}(k)}{b_{\mathrm{fid}}+f_{\mathrm{fid}}(\hat{\bs{k}}\cdot\hat{\bs{\eta}})^{2}}\text{,}
\end{equation}
where $f_\mathrm{fid}$ is the fiducial growth rate, $b_\mathrm{fid}$ is the fiducial linear galaxy bias,  $\hat{\bs{\eta}}$ is the line-of-sight and $\mathcal{S}$ is the smoothing kernel, whose functional form is given by
\begin{equation}
\mathcal{S}(\bs{k}) = \exp{\{-k^2\Sigma_s^2 / 2\}}\text{,}\label{eq:smoothing-kernel}
\end{equation}
with $\Sigma_s$ the smoothing scale.

The continuity equation implies
\begin{align}
n_d(\bs{x}) &= \int \dd^3x' n_g(\bs{x}') \delta_D(\bs{x} - \bs{x}' - \bs{s}(\bs{x}'))\\
n_s(\bs{x}) &= \int \dd^3x' n_r(\bs{x}') \delta_D(\bs{x} - \bs{x}' - \bs{s}(\bs{x}'))\text{.}
\end{align}
In Fourier space, they become
\begin{equation}
n_{d}(\bs{k})=\int\dd^{3}x\,n_{g}(\bs{x})e^{-i\bs{k}\cdot[\bs{x}+\bs{s}(\bs{x})]}\qquad n_{s}(\bs{k})=\int\dd^{3}x\,n_{r}(\bs{x})e^{-i\bs{k}\cdot[\bs{x}+\bs{s}(\bs{x})]}\text{.}
\end{equation}
Expanding the exponential term, we can then relate the reconstructed overdensity field $\delta_*$ with the galaxy and random overdensity field \citep{Shirasaki:2020vkk}
\begin{align}
\delta_{*}(\bs{k})	&=\frac{n_{d}(\bs{k})-\alpha n_{s}(\bs{k})}{\bar{n}_g}\\
&\equiv\int\dd^{3}x\,[\delta_{g}(\bs{x})-\delta_{r}(\bs{x})]e^{-i\bs{k}\cdot[\bs{x}+\bs{s}(\bs{x})]}\\
&=\sum_{n=0}^{\infty}\frac{1}{n!}\int_{\bs{k}=\bs{p}_{1\dots n+1}}[\bs{k}\cdot\bs{R}(\bs{p}_{2})]\cdots[\bs{k}\cdot\bs{R}(\bs{p}_{n+1})][\delta_{g}(\bs{p}_{1})-\delta_{r}(\bs{p}_{1})]\delta_{g}(\bs{p}_{2})\cdots\delta_{g}(\bs{p}_{n+1})\\
&\simeq\sum_{n=0}^{\infty}\frac{1}{n!}\int_{\bs{k}=\bs{p}_{1\dots n+1}}[\bs{k}\cdot\bs{R}(\bs{p}_{2})]\cdots[\bs{k}\cdot\bs{R}(\bs{p}_{n+1})]\delta_{g}(\bs{p}_{1})\delta_{g}(\bs{p}_{2})\cdots\delta_{g}(\bs{p}_{n+1})\\
&=\sum_{n=1}^{\infty}\int_{\bs{k}=\bs{p}_{1\dots n}}R_{n}(\bs{p}_{1},\cdots,\bs{p}_{n})\delta_{g}(\bs{p}_{1})\cdots\delta_{g}(\bs{p}_{n})\text{,}\label{eq:delta*}
\end{align}
where we use the notation
\begin{equation}
\int_{\bs{k}=\bs{p}_{1\dots n}} \equiv \int \frac{\dd^3 p_1}{(2\pi)^3}\cdots\int \frac{\dd^3 p_n}{(2\pi)^3} (2\pi)^3\delta_D(\bs{k}-\bs{p}_{1\dots n})\text{.}
\end{equation}
The overdensity of randoms is defined by $\delta_r(\bs{x}) \equiv \alpha n_r(\bs{x}) / \bar{n}_g - 1$, and has been neglected in the final expression. This is because the contribution from $\delta_r$ is generally $\alpha\sim 1/50$ times smaller compared to that from $\delta_g$. For example, in the shot noise dominated region we have $P^\NN_{rr} = \alpha P^\NN_{gg}$. The last line defines the $n$-th order symmetrized reconstruction kernel. The first $3$ terms are given by 
\begin{align}
R_{1}(\bs{p}_{1}) &=1\\
R_{2}(\bs{p}_{1},\bs{p}_{2}) &=\frac{1}{2}\bs{p}_{12}\cdot[\bs{R}(\bs{p}_{1})+\bs{R}(\bs{p}_{2})]\\
R_{3}(\bs{p}_{1},\bs{p}_{2},\bs{p}_{3})	&=\frac{1}{6}\left\{ [\bs{p}_{123}\cdot\bs{R}(\bs{p}_{1})][\bs{p}_{123}\cdot\bs{R}(\bs{p}_{2})]+(\text{2 cyc.})\right\} \text{.}
\end{align}

Analogous to the galaxy number density, the discreteness form of random number density is described by
\begin{equation}
n_r(\bs{x}) = \sum_i n_{r,i} \delta_D(\bs{x} - \bs{x}_i)\text{,}
\end{equation}
where $n_{r,i}$ is the random occupation number satisfying $n_{r,i} = n_{r,i}^2$. In addition, since the generation of random catalogues are independent of galaxies, galaxies and randoms are never in the same cell, i.e.,
\begin{equation}
n_{g,i} n_{r,i} = 0\text{.}
\end{equation}
$n_d, n_s$ and $n_*$ can then be expressed in discrete forms
\begin{align}
n_d(\bs{x}) &= \sum_i n_{g,i} \delta_D(\bs{x} - \bs{x}_i - \bs{s}_i) &n_d(\bs{k}) &= \sum_i n_{g,i} e^{-i\bs{k} \cdot (\bs{x}_i + \bs{s}_i)} \\
n_s(\bs{x}) &= \sum_i n_{r,i} \delta_D(\bs{x} - \bs{x}_i - \bs{s}_i) &n_s(\bs{k}) &= \sum_i n_{r,i} e^{-i\bs{k} \cdot (\bs{x}_i + \bs{s}_i)} \\
n_*(\bs{x}) &= \sum_i (n_{g,i} - \alpha n_{r,i}) \delta_D(\bs{x} - \bs{x}_i - \bs{s}_i) &n_*(\bs{k}) &= \sum_i (n_{g,i} - \alpha n_{r,i}) e^{-i\bs{k} \cdot (\bs{x}_i + \bs{s}_i)}\text{,}
\end{align}
with $\bs{s}_i$ the shift field evaluated at the position of the $i$-th cell. Using the notation
\begin{equation}
\delta_{*,i} = \frac{1}{\bar{n}_g} (n_{g,i} - \alpha n_{r,i}) e^{-i\bs{k} \cdot (\bs{x}_i + \bs{s}_i)}\text{,}
\end{equation}
the post-reconstruction power spectrum, bispectrum and trispectrum are defined as follows
\begin{align}
\sum_{i\not= j}  \langle \delta_{*,i}(\bs{k}_1) \delta_{*,j}(\bs{k}_2) \rangle_c &= V\delta^K_{\bs{k}_{12}} P_{**}(\bs{k}_1)\\
\sum_{i\not= j\not= k}  \langle \delta_{*,i}(\bs{k}_1) \delta_{*,j}(\bs{k}_2) \delta_{*,k}(\bs{k}_3) \rangle_c &= V\delta^K_{\bs{k}_{123}} B_{***}(\bs{k}_1, \bs{k}_2, \bs{k}_3)\\
\sum_{i\not= j\not= k\not= l}  \langle \delta_{*,i}(\bs{k}_1) \delta_{*,j}(\bs{k}_2) \delta_{*,k}(\bs{k}_3) \delta_{*,l}(\bs{k}_4) \rangle_c &= V\delta^K_{\bs{k}_{1234}} T_{****}(\bs{k}_1, \bs{k}_2, \bs{k}_3, \bs{k}_4)\text{.}
\end{align}
Their shot noise effect included counter parts and cross correlations between pre- and post-reconstruction density field can be very similarly defined.

\subsection{The shot noise effect after reconstruction}
Contrary to the pre-reconstruction case, the shot noise effect, even if Poissonian, cannot be removed by subtracting the self-pair contribution. \citet{Sugiyama:2024ggt} developed a method to compute the shot noise effect by counting self-pairs in the galaxy density field. We will follow this approach and derive the impact on bispectrum and trispectrum.

\begin{figure}
    \centering
    \includegraphics[width=1.0\linewidth]{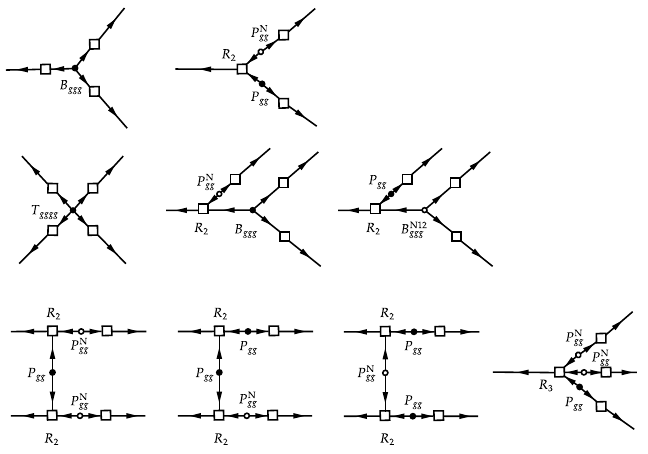}
    \caption{Diagram of post-reconstruction bispectrum $B_{***}$ and trispectrum $T_{****}$.}
    \label{fig:recon-diagram}
\end{figure}

We only need to expand $\delta_*$ up to the $3$rd order as we are only interested in the tree-level bispectrum and trispectrum
\begin{align}
\delta_{*}(\bs{k})&=\sum_{i}\delta_{g,i}(\bs{k})+\sum_{i,i'}\int_{\bs{k}=\bs{p}_{12}}R_{2}(\bs{p}_{1},\bs{p}_{2})\delta_{g,i}(\bs{p}_{1})\delta_{g,i'}(\bs{p}_{2})\nonumber\\&+\sum_{i,i',i''}\int_{\bs{k}=\bs{p}_{123}}R_{3}(\bs{p}_{1},\bs{p}_{2},\bs{p}_{3})\delta_{g,i}(\bs{p}_{1})\delta_{g,i'}(\bs{p}_{2})\delta_{g,i''}(\bs{p}_{3})+\mathcal{O}(\delta_{g}^{4})\text{.}
\end{align}
The calculation is then very similar to the pre-reconstruction galaxy bispectrum and trispectrum, except that we need to differentiate between the shot noise included and excluded correlators when connecting internal legs in the diagram. For example, the first line of Fig.~\ref{fig:recon-diagram} shows the diagram of $B_{***}$, where internal legs represent $\delta_g$ while external legs represent $\delta_*$. Filled circles denote galaxy correlators without the shot noise effect, while empty circles include the shot noise. Therefore, the post-reconstruction bispectrum is given by 
\begin{align}
B_{***}(\bs{k}_1, \bs{k}_2, \bs{k}_3) &= R_1(\bs{k}_1) R_1(\bs{k}_2) R_1(\bs{k}_3) B_{ggg} (\bs{k}_1, \bs{k}_2, \bs{k}_3) \nonumber\\
&+ \left\{ R_2(-\bs{k}_2, -\bs{k}_3) R_1(\bs{k}_2) R_1(\bs{k}_3)\left[P_{gg}(\bs{k}_2) P^\NN_{gg}(\bs{k}_3) + (\bs{k}_2\leftrightarrow\bs{k}_3)\right] + (2\text{ cyc.})\right\} \label{eq:B***} \text{.}
\end{align}
This expression recovers eq(22) in \citet{Shirasaki:2020vkk} if we rewrite $R_2$ in terms of the perturbation kernel $Z_2^\mathrm{rec}$ after reconstruction.

The cross correlation can be similarly obtained as
\begin{align}
B_{g**}(\bs{k}_1, \bs{k}_2, \bs{k}_3) &= R_1(\bs{k}_2) R_1(\bs{k}_3) B_{ggg} (\bs{k}_1, \bs{k}_2, \bs{k}_3) \nonumber \\
&+ \left\{ R_2(-\bs{k}_1, -\bs{k}_2) R_1(\bs{k}_2)\left[P_{gg}(\bs{k}_1) P^\NN_{gg}(\bs{k}_2) + (\bs{k}_1\leftrightarrow\bs{k}_2)\right] + (\bs{k}_2\leftrightarrow\bs{k}_3)\right\} \label{eq:Bg**}\\
B_{gg*}(\bs{k}_1, \bs{k}_2, \bs{k}_3) &= R_1(\bs{k}_3) B_{ggg}(\bs{k}_1, \bs{k}_2, \bs{k}_3) + R_2(-\bs{k}_1, -\bs{k}_2) R_1(\bs{k}_3)[P_{gg}(\bs{k}_1)P_{gg}^\NN(\bs{k}_2) + (\bs{k}_1 \leftrightarrow \bs{k}_2)] \text{.} \label{eq:Bgg*}
\end{align}

The $2$nd and the $3$rd line of Fig.~\ref{fig:recon-diagram} shows the diagram of the post-reconstruction trispectrum, which gives
\begin{align}
T_{****}(\bs{k}_{1},\bs{k}_{2},\bs{k}_{3},\bs{k}_{4})
&=R_1(\bs{k}_1) R_1(\bs{k}_2) R_1(\bs{k}_3) R_1(\bs{k}_4) T_{gggg}(\bs{k}_{1},\bs{k}_{2},\bs{k}_{3},\bs{k}_{4})\nonumber\\
&+R_{2}(-\bs{k}_{2},\bs{k}_{1}+\bs{k}_{2})R_{1}(\bs{k}_{2})R_{1}(\bs{k}_{3})R_{1}(\bs{k}_{4})\nonumber\\
&\phantom{+}\times[P_{gg}^{\NN}(\bs{k}_{2})B_{ggg}(\bs{k}_{3},\bs{k}_{4},\bs{k}_{1}+\bs{k}_{2})+P_{gg}(\bs{k}_{2})B_{ggg}^{\NN12}(\bs{k}_{3},\bs{k}_{4},\bs{k}_{1}+\bs{k}_{2})]\nonumber\\
&\phantom{+}+(11\text{ perm.})\nonumber\\
&+R_{2}(-\bs{k}_{4},\bs{k}_{1}+\bs{k}_{4})R_{2}(-\bs{k}_{3},\bs{k}_{2}+\bs{k}_{3})R_{1}(\bs{k}_{3})R_{1}(\bs{k}_{4})\nonumber\\
&\phantom{+}\times[P_{gg}^{\NN}(\bs{k}_{4})P_{gg}^{\NN}(\bs{k}_{3})P_{gg}(\bs{k}_{1}+\bs{k}_{4})+P_{gg}^{\NN}(\bs{k}_{4})P_{gg}(\bs{k}_{3})P_{gg}^{\NN}(\bs{k}_{1}+\bs{k}_{4})\nonumber\\
&\phantom{+}+P_{gg}(\bs{k}_{4})P_{gg}^{\NN}(\bs{k}_{3})P_{gg}^{\NN}(\bs{k}_{1}+\bs{k}_{4})+P_{gg}(\bs{k}_{4})P_{gg}(\bs{k}_{3})P_{gg}^{\NN}(\bs{k}_{1}+\bs{k}_{4})]\nonumber\\
&\phantom{+}+(11\text{ perm.})\nonumber\\
&+R_{3}(\bs{k}_{2},\bs{k}_{3},\bs{k}_{4})R_{1}(\bs{k}_{2})R_{1}(\bs{k}_{3})R_{1}(\bs{k}_{4})[2P_{gg}(\bs{k}_{2})P_{gg}^{\NN}(\bs{k}_{3})P_{gg}^{\NN}(\bs{k}_{4})+(2\text{ cyc.})]\nonumber\\
&\phantom{+}+(3\text{ cyc.})\text{.}
\end{align}
Similarly the cross correlation trispectrum with the full shot noise effect is given by 
\begin{align}
T_{gg**}^{\NN}(\bs{k}_{1},\bs{k}_{2},\bs{k}_{3},\bs{k}_{4})&=R_1(\bs{k}_3) R_1(\bs{k}_4) T_{gggg}^{\NN}(\bs{k}_{1},\bs{k}_{2},\bs{k}_{3},\bs{k}_{4})\nonumber\\
&+\left\{ \left[2R_{2}(-\bs{k}_{1},-\bs{k}_{23})P_{gg}^{\NN}(\bs{k}_{1})B_{ggg}^{\NN}(\bs{k}_{2},\bs{k}_{3},-\bs{k}_{23})+(2\text{ cyc.})\right]+(\bs{k}_{3}\leftrightarrow\bs{k}_{4})\right\} \nonumber\\
&+\left\{ 4R_{2}(-\bs{k}_{1},-\bs{k}_{23})R_{2}(-\bs{k}_{2},\bs{k}_{23})P_{gg}^{\NN}(\bs{k}_{1})P_{gg}^{\NN}(\bs{k}_{2})P_{gg}^{\NN}(\bs{k}_{23})+(\bs{k}_{1}\leftrightarrow\bs{k}_{2})\right\} \nonumber\\
&+\left\{ 6R_{3}(-\bs{k}_{1},-\bs{k}_{2},-\bs{k}_{3})P_{gg}^{\NN}(\bs{k}_{1})P_{gg}^{\NN}(\bs{k}_{2})P_{gg}^{\NN}(\bs{k}_{3})+(\bs{k}_{3}\leftrightarrow\bs{k}_{4})\right\} \text{.} \label{eq:TNgg**}
\end{align}

Note that the above equations only hold perturbatively, and one should expand $P_{gg}, B_{ggg}$ and $T_{gggg}$ at the tree-level. Although in this work we only assume the Poissionian shot noise, the full stochastic term \citep{Perko:2016puo} can in principle to be plugged into the above equations or re-parameterized. We refer the readers to \citet{Ebina:2024zkv} for a relevant discussion.

\subsection{The Gaussian covariance}
The pre- and post-reconstruction power spectrum estimators for an individual mode is given by 
\begin{align}
\widehat{P}_{gg}(\bs{k}) &= \frac{1}{V\bar{n}_g^2}\left\{ |n_g(\bs{k})|^2 - N_g \right\} \label{eq:Pgg-estimator} \\ 
\widehat{P}_{**}(\bs{k}) &= \frac{1}{V\bar{n}_g^2}\left\{ |n_*(\bs{k})|^2 - (1+\alpha)N_g \right\} \label{eq:P**-estimator}\text{,}
\end{align}
where $N_g$ is the total number of galaxies. These two estimators are shot noise subtracted, and can be expressed in terms of $\delta_{g,i}$ and $\delta_{*,i}$
\begin{equation}
\widehat{P}_{gg}(\bs{k}) = \frac{1}{V}\sum_{i\not=j} \delta_{g,i}(\bs{k}) \delta_{g,j}(-\bs{k}) \qquad \widehat{P}_{**}(\bs{k}) = \frac{1}{V}\sum_{i\not=j} \delta_{*,i}(\bs{k}) \delta_{*,j}(-\bs{k})\text{.}
\end{equation}
The cross power spectrum estimator between pre- and post-reconstruction density field is written as
\begin{equation}
\widehat{P}^\NN_{g*}(\bs{k}) = \frac{1}{V\bar{n}_g^2} n_g(\bs{k}) n_*(\bs{k}) = \frac{1}{V} \sum_{i,j} \delta_{g, i}(\bs{k}) \delta_{*, i}(-\bs{k})\text{,} \label{eq:Pg*-estimator}
\end{equation}
here we do not subtract the shot noise.

The pre-reconstruction covariance is then given by\footnote{Terms involving $\langle \delta_{g,i}(\bs{k})\rangle_c$ vanish because $k_1\not=0$ and $k_2\not=0$.}
\begin{align}
C_{gggg}(\bs{k}_1, \bs{k}_2) &= \langle \widehat{P}_{gg}(\bs{k}_1) \widehat{P}_{gg}(\bs{k}_2) \rangle - \langle \widehat{P}_{gg}(\bs{k}_1) \rangle \langle \widehat{P}_{gg}(\bs{k}_2) \rangle \\
&=\frac{1}{V^2} \sum_{i,k} \langle \delta_{g,i}(\bs{k}_1) \delta_{g,k}(\bs{k}_2) \rangle_c \sum_{j,l} \langle \delta_{g,j}(-\bs{k}_1) \delta_{g,l}(-\bs{k}_2) \rangle_c + (\bs{k}_2\leftrightarrow -\bs{k}_2) \nonumber\\
&+ \frac{1}{V^2} \sum_{i\not=j, k\not=l} \langle \delta_{g,i}(\bs{k}_1) \delta_{g,j}(-\bs{k}_1) \delta_{g,k}(\bs{k}_2) \delta_{g,l}(-\bs{k}_2) \rangle_c \\
&\equiv C_{gggg}^\mathrm{G} (\bs{k}_1, \bs{k}_2) + C_{gggg}^\mathrm{T} (\bs{k}_1, \bs{k}_2)\text{.}
\end{align}
The last line defines the Gaussian and non-Gaussian contribution. The Gaussian contribution evaluates
\begin{equation}
C_{gggg}^\mathrm{G}(\bs{k}_1, \bs{k}_2) = [P_{gg}^\NN(\bs{k}_1)]^2(\delta^K_{\bs{k}_1 + \bs{k}_2} + \delta^K_{\bs{k}_1 - \bs{k}_2})\text{.}
\end{equation}

The post-reconstruction and the cross covariance between pre- and post-reconstruction covariance are similarly given as 
\begin{align}
C_{****}^\mathrm{G}(\bs{k}_1, \bs{k}_2) &= [P_{**}^\NN(\bs{k}_1)]^2(\delta^K_{\bs{k}_1 + \bs{k}_2} + \delta^K_{\bs{k}_1 - \bs{k}_2})\\
C_{gg**}^\mathrm{G}(\bs{k}_1, \bs{k}_2) &= [P_{g*}^\NN(\bs{k}_1)]^2(\delta^K_{\bs{k}_1 + \bs{k}_2} + \delta^K_{\bs{k}_1 - \bs{k}_2})\text{.}
\end{align}

\subsection{The non-Gaussian covariance}
The non-Gaussian covariance for pre-reconstruction is given by \citet{Wadekar:2019rdu, Sugiyama:2019ike}. The post-reconstruction covariance has the same structure as the pre-reconstruction one
\begin{align}
C_{****}^{\mathrm{T}}(\bs{k}_{1},\bs{k}_{2})&=\frac{1}{V}T_{****}(\bs{k}_{1},-\bs{k}_{1},\bs{k}_{2},-\bs{k}_{2})\nonumber\\
&+\frac{1}{V}\frac{1}{\bar{n}_{g}}\left[B_{***}(\bs{k}_{12},-\bs{k}_{1},-\bs{k}_{2})+B_{***}(\bs{k}_{1}-\bs{k}_{2},-\bs{k}_{1},\bs{k}_{2})+(\bs{k}_{i}\leftrightarrow-\bs{k}_{i})\right]\nonumber\\
&+\frac{1}{V}\frac{1}{\bar{n}_{g}^{2}}[P_{**}(\bs{k}_{1}+\bs{k}_{2})+P_{**}(\bs{k}_{1}-\bs{k}_{2})]\text{.}
\end{align}

Let us now consider the non-Gaussian covariance between pre- and post-reconstruction power spectra
\begin{equation}
C_{gg**}^\mathrm{T} (\bs{k}_1, \bs{k}_2) = \frac{1}{V^2}\sum_{i\not=j, k\not=l} \langle \delta_{g,i}(\bs{k}_1) \delta_{g,j}(-\bs{k}_1) \delta_{*, k}(\bs{k}_2) \delta_{*,l}(-\bs{k}_2) \rangle_c\text{.}
\end{equation}
Since terms like
\begin{equation}
\sum_{i\not=j\not=l, i=k}\cdots = \sum_{i\not=j\not=l} \frac{1}{\bar{n}_g^2} n_{g,i} e^{-i\bs{k}_{12}\cdot \bs{x}_i} e^{-i\bs{k}_2\cdot{s}_i} \delta_{g,j}(-\bs{k}_1)\delta_{*,l}(-\bs{k}_2)
\end{equation}
cannot be expressed in terms of $\delta_g$ and $\delta_*$, we instead expand the summation as follows \citep[see eq(17), eq(18) and eq(19) in ][]{Sugiyama:2019ike}
\begin{equation}
\sum_{i\not=j, k\not=l} =  \sum_{i,j,k,l} - \sum_{i=j, k, l} - \sum_{k=l, i, j} + \sum_{i=j, k=l}\text{.}
\end{equation}
The first term is
\begin{equation}
\sum_{i,j,k,l} \cdots  = \frac{1}{V} T_{gg**}^\NN (\bs{k}_1, -\bs{k}_1, \bs{k}_2, -\bs{k}_2)\text{.}
\end{equation}
Using eq(\ref{eq:delta_gi_contract}) and
\begin{equation}
\delta_{*,i}(\bs{0}) \simeq \delta_{g,i}(\bs{0}) \qquad \delta_{*,i}(\bs{k}_1) \delta_{*,i}(\bs{k}_2) \simeq \frac{1}{\bar{n}_g} \delta_{*,i}(\bs{k}_{12}) \text{,}
\end{equation}
the second term expands
\begin{align}
\sum_{i=j,k,l}\cdots &= \sum_{i\not=k\not=l} + \sum_{i=k, k\not=l} +  \sum_{i=l, i\not=k} + \sum_{k=l, i\not=k} + \sum_{i=j=k} \\
&=\frac{1}{V}\frac{1}{\bar{n}_g} \left\{ B_{g**}(\bs{0}, \bs{k}_2, -\bs{k}_2) + \frac{2}{\bar{n}_g} P_{**}(\bs{k}_2) + \frac{1}{\bar{n}_g} P_{g*}(\bs{0}) + \frac{1}{\bar{n}_g^2} \right\}\text{.}
\end{align}
The third term expands
\begin{equation}
\sum_{k=l, i,j} \cdots = \frac{1}{V}\frac{1}{\bar{n}_g} \left\{ B_{gg*}(\bs{k}_1, -\bs{k}_1, \bs{0}) + \frac{2}{\bar{n}_g} P_{gg}(\bs{k}_2) + \frac{1}{\bar{n}_g} P_{g*}(\bs{0}) + \frac{1}{\bar{n}_g^2} \right\}\text{,}
\end{equation}
and the last term is given by 
\begin{equation}
\sum_{i=j, k=l} \cdots = \frac{1}{V}\frac{1}{\bar{n}_g^2} \left\{ P_{g*}(\bs{0}) + \frac{1}{\bar{n}_g} \right\}\text{.}
\end{equation}
Therefore, we obtain
\begin{align}
C_{gg**}^\mathrm{T} (\bs{k}_1, \bs{k}_2) &= \frac{1}{V}T^\NN_{gg**}(\bs{k}_1, -\bs{k}_1, \bs{k}_2, -\bs{k}_2) \nonumber \\
&- \frac{1}{V}\frac{1}{\bar{n}_g}\left\{ B_{gg*}(\bs{k}_{1},-\bs{k}_{1},\bs{0})+B_{g**}(\bs{0},\bs{k}_{2},-\bs{k}_{2})+\frac{2}{\bar{n}_{g}}P_{gg}(\bs{k}_{1})+\frac{2}{\bar{n}_{g}}P_{**}(\bs{k}_{2})+\frac{1}{\bar{n}_{g}}P_{g*}(\bs{0})+\frac{1}{\bar{n}_{g}^{2}}\right\} \text{.}
\end{align}
Using the fact that at the tree level
\begin{equation}
P_{gg}(\bs{k})=P_{**}(\bs{k}) \qquad\qquad P_{g*}(\bs{0}) = P_{gg}(\bs{0})\text{,}
\end{equation}and substituting eq(\ref{eq:Bg**}), eq(\ref{eq:Bgg*}) and eq(\ref{eq:TNgg**}) into the above formula, the final expression writes
\begin{align}
C_{gg**}^{\mathrm{T}}(\bs{k}_{1},\bs{k}_{2})&=C_{gggg}^{\mathrm{T}}(\bs{k}_{1},\bs{k}_{2})\nonumber\\
&+\frac{1}{V}\left\{ \left[2R_{2}(-\bs{k}_{1},\bs{k}_{1}-\bs{k}_{2})P_{gg}^{\NN}(\bs{k}_{1})B_{ggg}^{\NN}(-\bs{k}_{1},\bs{k}_{2},\bs{k}_{1}-\bs{k}_{2})+(\bs{k}_{1}\leftrightarrow-\bs{k}_{1})\right]+(\bs{k}_{2}\leftrightarrow-\bs{k}_{2})\right\} \nonumber\\
&+\frac{1}{V}\left\{ 4R_{2}(-\bs{k}_{1},\bs{k}_{1}-\bs{k}_{2})^{2}P_{gg}^{\NN}(\bs{k}_{1})^{2}P_{gg}^{\NN}(-\bs{k}_{1}+\bs{k}_{2})+(\bs{k}_{1}\leftrightarrow-\bs{k}_{1})\right\}\nonumber\\
&+\frac{1}{V}\left\{ 6R_{3}(-\bs{k}_{1},\bs{k}_{1},-\bs{k}_{2})P_{gg}^{\NN}(\bs{k}_{1})^{2}P_{gg}^{\NN}(\bs{k}_{2})+(\bs{k}_{2}\leftrightarrow-\bs{k}_{2})\right\} \nonumber\\
&+\frac{1}{V}\left\{ 2R_{2}(-\bs{k}_{2},\bs{0})P_{gg}^{\NN}(\bs{k}_{2})B_{ggg}^{\NN}(\bs{k}_{1},-\bs{k}_{1},\bs{0})+(\bs{k}_{2}\leftrightarrow-\bs{k}_{2})\right\} \nonumber\\
&-\frac{1}{V}\frac{2}{\bar{n}_{g}}\left\{ R_{2}(\bs{0},\bs{k}_{2})P_{gg}(\bs{0})P_{gg}^{\NN}(\bs{k}_{2})+R_{2}(\bs{0},\bs{k}_{2})P_{gg}^{\NN}(\bs{0})P_{gg}(\bs{k}_{2})\right\}\text{.}
\end{align}
The last two lines vanish in the case when there are no modes larger than the periodic simulation box we are considering.

\subsection{Numerical implementation}
In previous sections, we derived the covariance matrix of power spectrum estimators for a single mode. In the analysis, we will use the binned power spectrum projected in the Legendre basis, i.e., the power spectrum multipoles, namely,
\begin{equation}
\widehat{P}_\ell (k) = (2\ell + 1) \sum_{\bar{\bs{k}}} \widehat{P}(\bs{k}) \mathcal{L}_{\ell} (\hat{\bs{k}}\cdot\hat{\bs{\eta}})\text{,} \label{eq:multipole-estimator}
\end{equation}
where $\bs{k}$ takes discrete values with the unit of fundamental frequency $k_f=2\pi / L$, and $\mathcal{L}_\ell$ is the $\ell$-th order Legendre polynomial. Symbol $\sum_{\bar{\bs{k}}}$ represents the average of $\bs{k}$-modes inside a spherical shell with the wavenumber $|\bs{k}|$ satisfying $|\bs{k}|\in[k - \Delta k/2, k + \Delta k / 2)$ with $\Delta k$ the width of $k$-bins. This definition can be expressed in terms of a top-hat function $\Xi_k$
\begin{equation}
\sum_{\bar{\bs{k}}}\equiv\sum_{\bs{k}'}\Xi_{k}(\bs{k}')\qquad\Xi_{k}(\bs{k}')\equiv\begin{cases}
\frac{1}{N_{k}} & \text{if }|\bs{k}'|\in[k-\Delta k/2,k+\Delta k/2)\\
0 & \text{otherwise}
\end{cases}\text{,}
\end{equation}
with $N_k$ the number of modes inside the spherical shell $N_k \simeq V_k / k_f^3$. $\Xi_{k}$ is an even function and has the following properties
\begin{align}
\sum_{\bar{\bs{k}}_{2}}g(\bs{k}_{1},\bs{k}_{2})\delta_{\bs{k}_{12}}^{K}	&=g(\bs{k}_{1},-\bs{k}_{1})\Xi_{k_{2}}(\bs{k}_{1})\\
\sum_{\bar{\bs{k}}_{1},\bar{\bs{k}}_{2}}g(\bs{k}_{1},\bs{k}_{2})\delta_{\bs{k}_{12}}^{K}	&=\frac{\delta_{k_{1},k_{2}}^{K}}{N_{k_{1}}}\sum_{\bar{\bs{k}}_{1}}g(\bs{k}_{1},-\bs{k}_{1})\text{,}
\end{align}
where $g$ is an arbitrary function, and the $2$nd equality assumes non-overlapping bins for $k_{1}$ and $k_{2}$. The two-argument Kronecker symbol $\delta^K_{k_1,k_2}$ is equal to $1$ when $k_1=k_2$ and vanishes otherwise.

The Gaussian covariance matrix of power spectrum multipoles becomes
\begin{equation}
C_{\ell_{1}\ell_{2}}^{\mathrm{G}}(k_{1},k_{2})=\delta_{k_{1},k_{2}}^{K}\frac{2}{N_{k_{1}}}(2\ell_{1}+1)(2\ell_{2}+1)\sum_{\bar{\bs{k}}_{1}}P^2(\bs{k}_{1})\mathcal{L}_{\ell_{1}}(\hat{\bs{k}}_{1}\cdot\hat{\bs{\eta}})\mathcal{L}_{\ell_{2}}(\hat{\bs{k}}_{1}\cdot\hat{\bs{\eta}})\text{,}
\end{equation}
where $P$ takes $P^\NN_{gg}$, $P^\NN_{**}$ and $P^\NN_{g*}$ for the pre-, post- and cross-correlation covariance matrix, respectively. The non-Gaussian part is given by 
\begin{equation}
C_{\ell_{1}\ell_{2}}^{\mathrm{T}}(k_{1},k_{2})=(2\ell_{1}+1)(2\ell_{2}+1)\sum_{\bar{\bs{k}}_{1},\bar{\bs{k}}_{2}}C^{\mathrm{T}}(\bs{k}_{1},\bs{k}_{2})\mathcal{L}_{\ell_{1}}(\hat{\bs{k}}_{1}\cdot\hat{\bs{\eta}})\mathcal{L}_{\ell_{2}}(\hat{\bs{k}}_{2}\cdot\hat{\bs{\eta}})\text{.}
\end{equation}

Although the above two equations accurately capture the binning effect in the power spectrum measurement, we will adopt the continuous limit and the thin shell approximation to reduce the computational cost, i.e.,
\begin{equation}
\sum_{\bar{\bs{k}}}\to\int_{\bar{\bs{k}}}\to\int_{\hat{\bs{k}}}\equiv\int\frac{\dd^{2}\hat{k}}{4\pi}\text{.}
\end{equation}
Therefore the Gaussian covariance matrix is given by
\begin{equation}
C_{\ell_{1}\ell_{2}}^{\mathrm{G}}(k_{1},k_{2})=\delta_{k_{1},k_{2}}^{K}\frac{2}{N_{k_{1}}}(2\ell_{1}+1)(2\ell_{2}+1)\int_{\hat{\bs{k}}_{1}}P^2(\bs{k}_{1})\mathcal{L}_{\ell_{1}}(\hat{\bs{k}}_{1}\cdot\hat{\bs{\eta}})\mathcal{L}_{\ell_{2}}(\hat{\bs{k}}_{1}\cdot\hat{\bs{\eta}})\text{.}
\end{equation}
In the calculation, we use the non-perturbative $P(\bs{k}_1)$ constructed by averaging over $\widehat{P}_0, \widehat{P}_2$ and $\widehat{P}_4$ measured from simulations. 
We note that we do not remove the binning effect for simplicity.

According to the rotational symmetry, the non-Gaussian part reduces to a $3$-dimensional integration
\begin{equation}
C_{\ell_{1}\ell_{2}}^{\mathrm{T}}(k_{1},k_{2}) = \frac{(2\ell_{1}+1)(2\ell_{2}+1)}{8\pi}\int_{-1}^{1}\dd\mu_{1}\int_{-1}^{1}\dd\mu_{2}\int_{0}^{2\pi}\dd\phi\,C^{\mathrm{T}}(k_{1},k_{2},\mu_{1},\mu_{2},\phi)\mathcal{L}_{\ell_{1}}(\mu_{1})\mathcal{L}_{\ell_{2}}(\mu_{2})\text{.}
\end{equation}

\section{Comparison with simulations}
\label{sec:dem}
To test the analytic covariance derived in the last section, we measure the numerical covariance matrix of power spectra multipoles from $15000$ QUIJOTE halo catalogues \footnote{\href{https://quijote-simulations.readthedocs.io/en/latest/halos.html}{https://quijote-simulations.readthedocs.io/en/latest/halos.html}} at redshift $z=0.5$. Each simulation evolves $512^3$ particles inside a periodic box of $1\,h^{-1}\,\mathrm{Gpc}$ length with the initial condition set by $2$LPT at $z=127$ and the fiducial cosmology $\{ \Omega_m, \Omega_b, h, n_s, \sigma_8, M_\nu\} = \{ 0.3175, 0.049, 0.6711, 0.9624, 0.834, 0 \}$. Dark matter halos are identified using the FoF algorithm \citep{Davis:1985rj}. In this work, we only consider haloes with a mass range of $10^{13.1-14.0}\,h^{-1}M_\odot$, leading to the mean halo number density of $\bar{n}_h \simeq 2.918\times 10^{-4}\,[h^{-1}\mathrm{Mpc}]^{-3}$. The linear bias is found to be $b_1\simeq 1.85$, obtained by running fits to the power spectrum multipoles with the \code{Velocileptors} EPT code \citep{Chen:2020zjt, Maus:2024dzi}.

\begin{figure}
    \centering
    \includegraphics[width=0.7\linewidth]{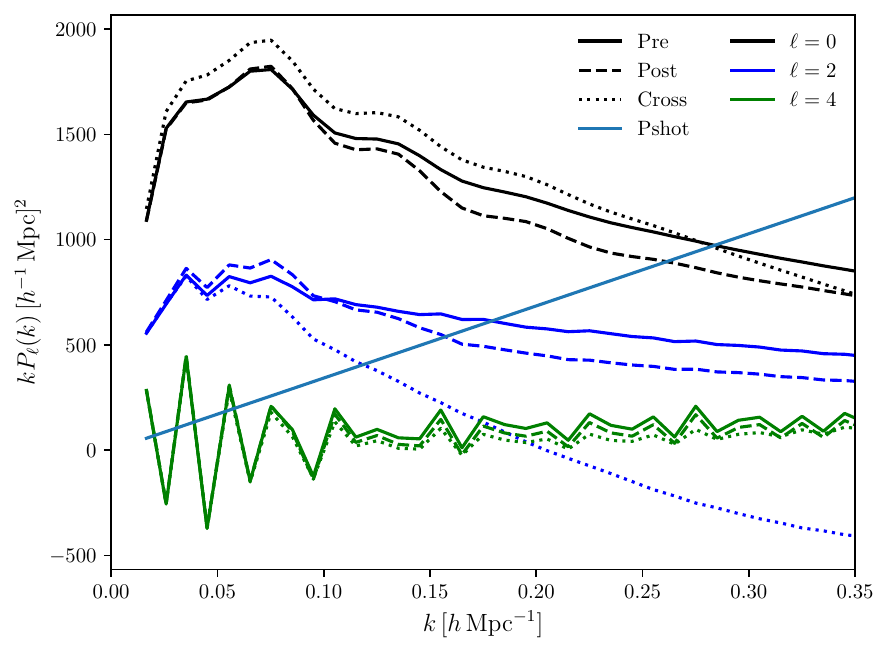}
    \caption{Solid lines, dashed lines and dotted lines show the measured pre- and post-recon power spectrum, and the cross power spectrum of our baseline, respectively. In this plot, the shot noise is removed from both the pre- and post-reconstruction power spectra, but is retained in the cross power spectrum.} 
    \label{fig:baseline-pk}
\end{figure}

We use the public code \code{pyrecon}\footnote{\href{https://github.com/cosmodesi/pyrecon}{https://github.com/cosmodesi/pyrecon}} and \code{pypower}\footnote{https://github.com/cosmodesi/pypower} to perform the density field reconstruction and power spectrum measurement \citep{Hand:2017irw}, respectively. The reconstruction algorithm is summarized as the following steps
\begin{enumerate}
    \item Interpolate halo particles on a $1000^3$ mesh with a box size of $1000\,h^{-1}\mathrm{Mpc}$ using the Cloud-In-Cell (CIC) scheme, then compute the overdensity and transform to Fourier space.
    \item Smooth the overdensity field using the kernel defined in eq(\ref{eq:smoothing-kernel}) with $\Sigma_s = 10, 15$ and $ 20\,h^{-1}\mathrm{Mpc}$.
    \item Assuming a fiducial value for the linear bias $b_\mathrm{fid} = b_1$ and the growth rate $f_\mathrm{fid}=f_\mathrm{true} = 0.7628$ or\footnote{The choice of $f_\mathrm{fid}=0$ is usually adopted in perturbation theory calculations \citep[e.g., ][]{Hikage:2019ihj,Sugiyama:2024eye} as it is not straightforward to obtain the analytic expression of the full $3$D power spectrum expanded in Legendre polynomials when $\hat{\bs{k}}\cdot\hat{\bs{\eta}}$ appears in the denominator.} $f_\mathrm{fid} = 0$, estimate the shift field on the grid by solving eq(\ref{eq:shift-field}).
    \item Displace halo particles with the estimated shift field interpolated at the particle position with the CIC scheme to obtain the displaced data catalogue.
    \item Generate a random catalogue with the number of randoms $50$ times the number of haloes. Random particles are then displaced in the same manner as haloes to obtain the displaced random catalogue.
\end{enumerate}
The baseline reconstruction setting we use is $b_\mathrm{fid}=b_1, \Sigma_s = 15\,h^{-1}\,\mathrm{Mpc}$ and $f_\mathrm{fid} = 0.7628$. The density field is then obtained by assigning particles on a $1000^3$ mesh using the Triangular Shaped Cloud (TSC) scheme with the window function correction applied \citep{Jing:2004fq}. When painting particles on the mesh, we additionally use $2$ interlaced grids \citep{Sefusatti:2015aex} shifted by $1/3$ and $2/3$ mesh cell size, respectively, to reduce the aliasing effect. Then, the pre-, post- and cross-power spectrum multipoles are estimated following eq(\ref{eq:Pgg-estimator}), eq(\ref{eq:P**-estimator}), eq(\ref{eq:Pg*-estimator}) and eq(\ref{eq:multipole-estimator}). In our analysis, we consider power spectrum measured between $k_\mathrm{min}=0.01\,h\,\mathrm{Mpc}^{-1}$ and $k_\mathrm{max}=0.35\,h\,\mathrm{Mpc}^{-1}$, with the $k$-bin size set by $\Delta k=0.01\,h\,\mathrm{Mpc}^{-1}$. The measurements are shown in Fig.~\ref{fig:baseline-pk}.

The derivations presented in the last section have assumed $\bar{n}_g$ is a fixed number in the denominator. It is, however, not true because the number of haloes varies between different realisations. So we have two choices in the power spectrum estimation. One choice is to normalize the estimator using the number density measured in each realization, or the ``local number density''. Another choice is to use the ensemble averaged number density, or the ``global number density''. We found these two choices lead to almost the same mean power spectra but different numerical covariances especially at the small scale, with the ``local number density'' case showing smaller correlations in both the diagonal elements and the non-diagonal parts. This effect is very similar to the local average effect as pointed out by \citet{dePutter:2011ah}, but their physical origins are different. The later is caused by the matter fluctuations at the survey scale, while in our case, the difference in the halo number density is caused by different initial conditions. On the other hand, comparing the analytic covariance with the ``global number density'' case is also not very consistent on the theory side, because it means the overdensity we defined can have fluctuations at the box scale. In any case, we will present both the ``local number density'' and the ``global number density'' measurements when comparing the theory prediction to simulations. The variation of number densities additionally leads to the variation of shot noise, but this effect has already been included in the derivation \citep[see discussions in][]{Smith:2008ut, Chan:2016ehg, Wadekar:2019rdu, Sugiyama:2019ike}.

\begin{figure}
\centering
\includegraphics[width=1.0\linewidth]{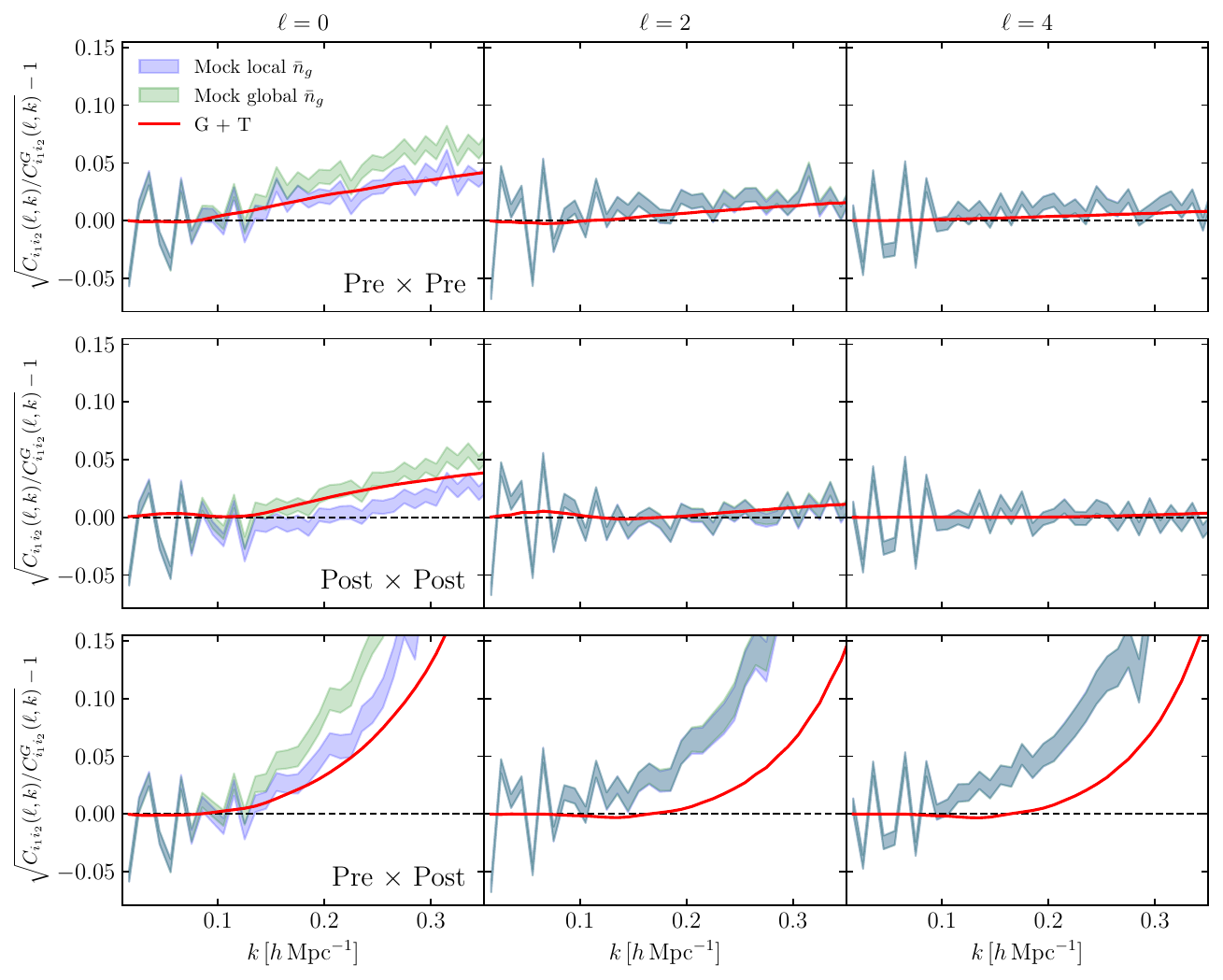}
\caption{The relative difference in the square root of diagonal covariance elements compared with the Gaussian prediction. Red lines show the result predicted by the analytic covariance matrix with the non-Gaussian contribution. Shaded bands show the numerical covariance matrix %\YW{
computed from $15000$ simulations
. 
}
\label{fig:baseline_diag}
\end{figure}

The full covariance is represented by $C_{i_1 i_2 \ell_1 \ell_2}(k_1, k_2)$, where $i_1$ and $i_2$ label either the pre- or post-reconstruction power spectrum. We will use symbol $C_{i_1 i_2}(\ell, k)$ to denote the diagonal part, i.e., $\ell_1=\ell_2=\ell$ and $k_1=k_2=k$
\begin{equation}
C_{i_1 i_2}(\ell, k) \equiv C_{i_1 i_2 \ell \ell}(k, k)\text{.}
\end{equation}
Fig.~\ref{fig:baseline_diag} compares the diagonal part of covariance matrices between numerical covariances and analytic covariances using the baseline reconstruction setting. The $1$-$\sigma$ error band is estimated by bootstrapping. As shown in the plot, the departure from the Gaussian prediction is around $5$ per cent\footnote{This argument depends on the size of $k$-bins $\Delta k$.} up to $k=0.35\,h\,\mathrm{Mpc}^{-1}$ for the auto-correlation covariance. The post-reconstruction covariance agrees with the Gaussian prediction better compared to the pre-reconstruction\footnote{We remind the readers the Gaussian contribution is computed in the non-perturbative way, so any deviations show the pure trispectrum effect.}. This is expected, because reconstruction undoes the nonlinear gravitational evolution, so that the induced correlations between measured power spectra are suppressed \citep{Hikage:2020fte}. However, the cross-correlation between pre- and post-reconstruction power spectra is much stronger compared to the Gaussian prediction, and the deviation exceeds $15$ per cent when $k$ goes beyond $0.25\,h\,\mathrm{Mpc}^{-1}$. Adding the non-Gaussian contribution to the analytic covariance improves the agreement between the theoretical prediction and the numerical covariance especially in the auto-correlation covariance, while the theory still under-predicts the cross-correlation covariance.

\begin{figure}
\centering
\includegraphics[width=1.0\linewidth]{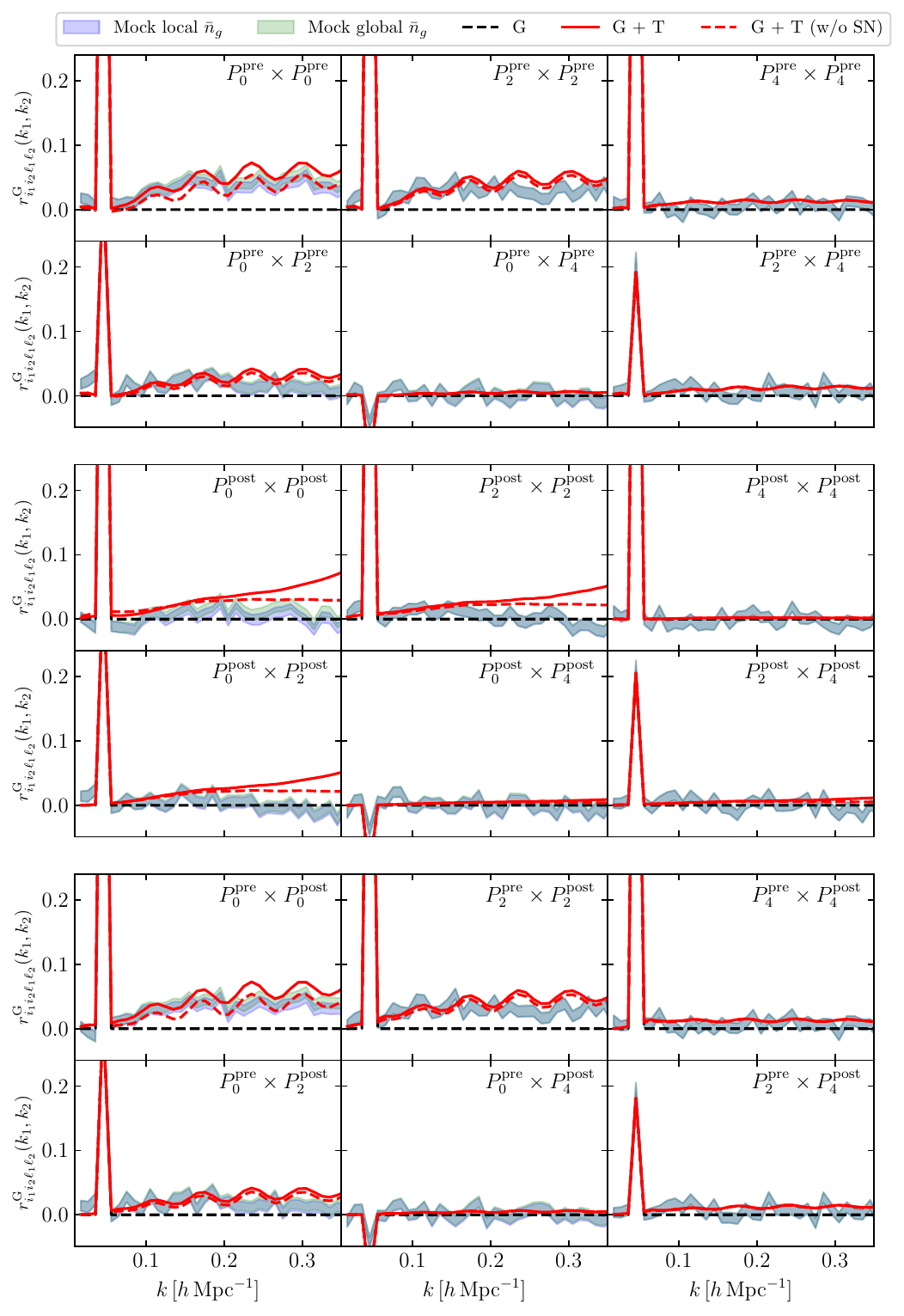}
\caption{Covariance matrix normalized by the diagonal part of the Gaussian covariance with $k_2=0.045\,h\,\mathrm{Mpc}^{-1}$. Black dashed lines show the Gaussian analytic covariance. Red solid lines show the non-Gaussian analytic covariance. Red dashed lines show the non-Gaussian analytic covariance but excluding the shot noise contribution when calculating the $C^\mathrm{T}$ term. Blue and Green bands are numerical covariance matrices estimated from $15000$ realizations.}
\label{fig:baseline_rcovG_0.045}
\end{figure}

\begin{figure}
\centering
\includegraphics[width=1.0\linewidth]{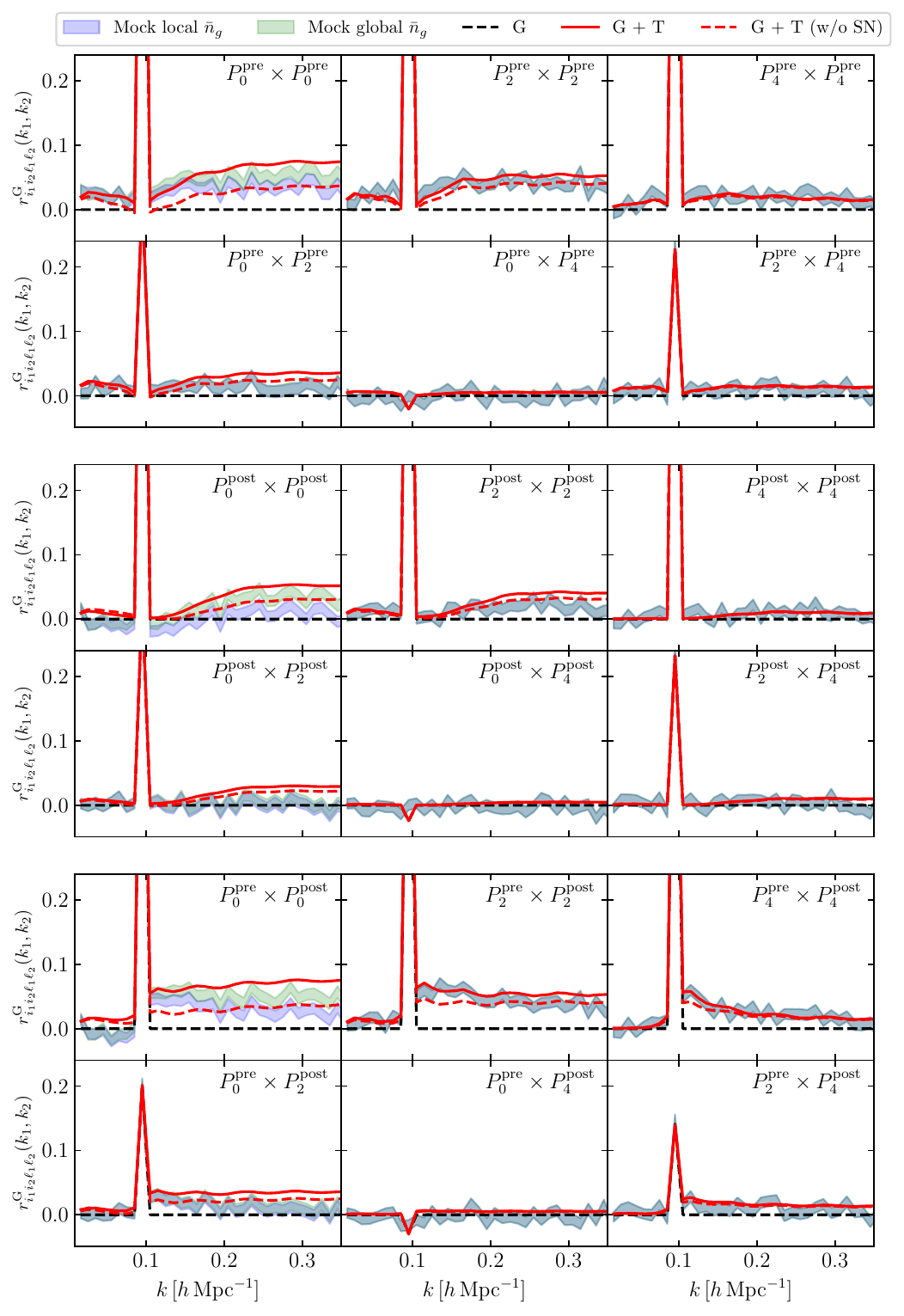}
\caption{Similar to Fig.~\ref{fig:baseline_rcovG_0.045}, but with $k_2=0.095\,h\,\mathrm{Mpc}^{-1}$.}
\label{fig:baseline_rcovG_0.095}
\end{figure}

\begin{figure}
\centering
\includegraphics[width=1.0\linewidth]{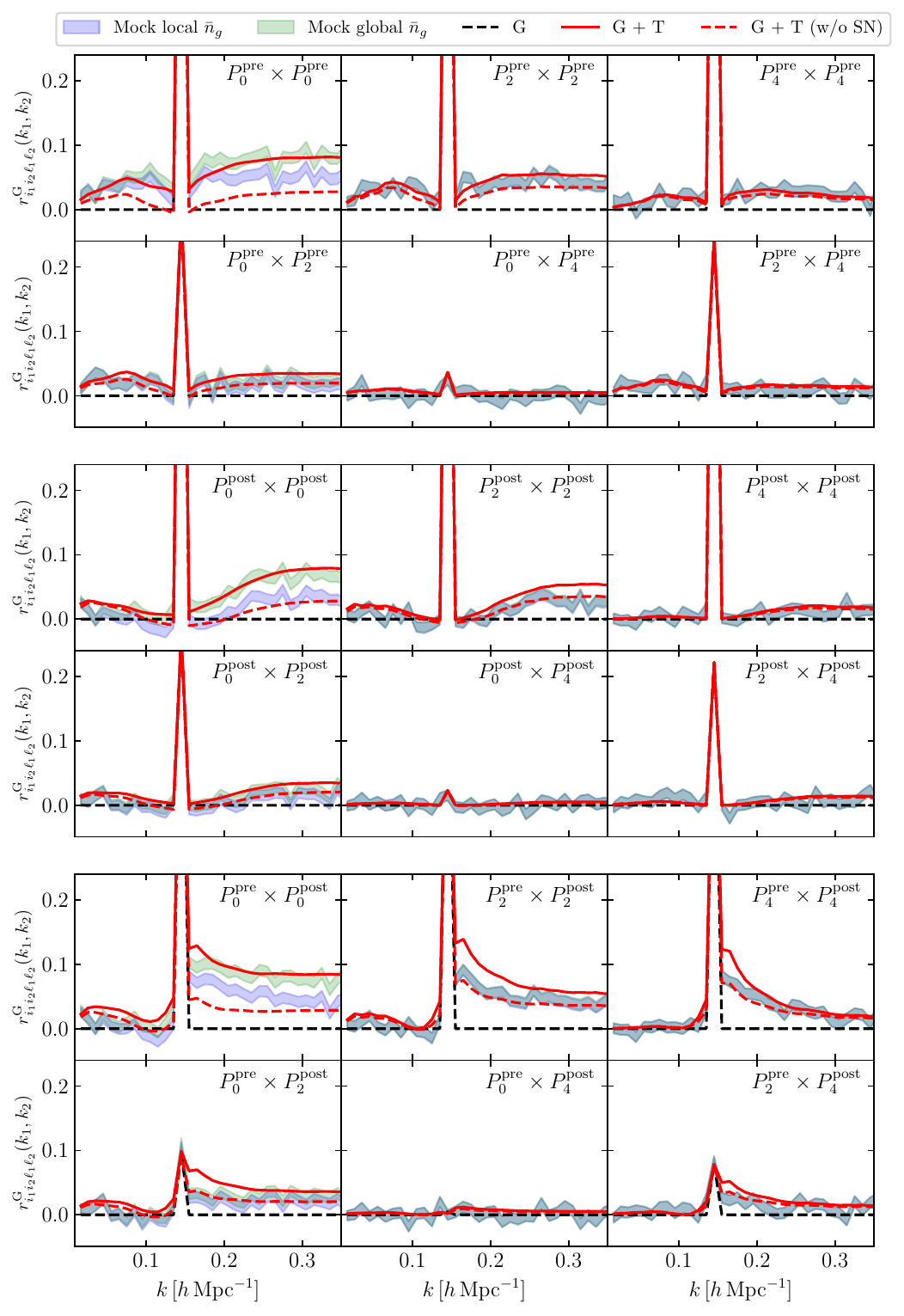}
\caption{Similar to Fig.~\ref{fig:baseline_rcovG_0.045}, but with $k_2=0.145\,h\,\mathrm{Mpc}^{-1}$.}
\label{fig:baseline_rcovG_0.145}
\end{figure}

\begin{figure}
\centering
\includegraphics[width=1.0\linewidth]{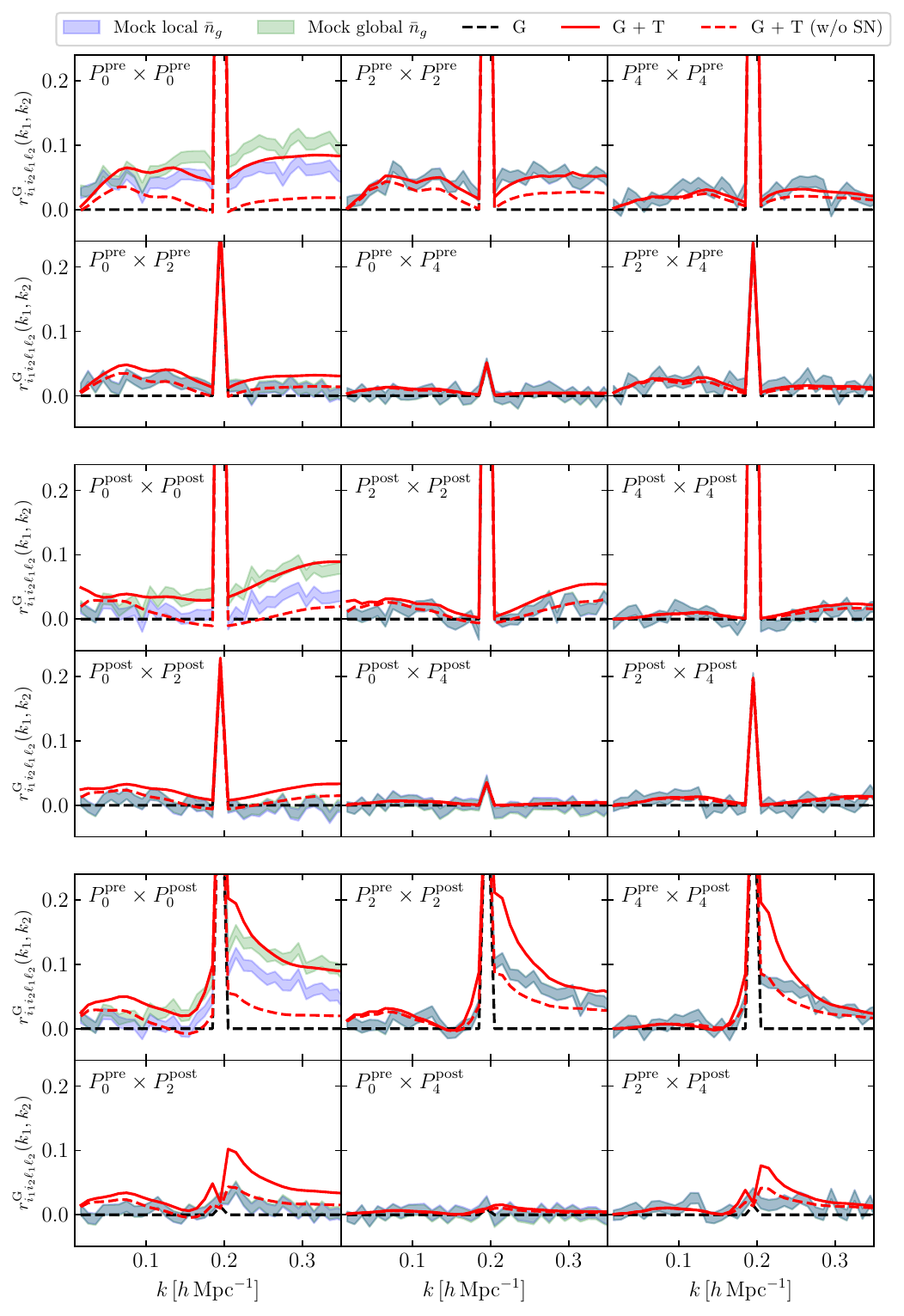}
\caption{Similar to Fig.~\ref{fig:baseline_rcovG_0.045}, but with $k_2=0.195\,h\,\mathrm{Mpc}^{-1}$.}
\label{fig:baseline_rcovG_0.195}
\end{figure}

\begin{figure}
\centering
\includegraphics[width=1.0\linewidth]{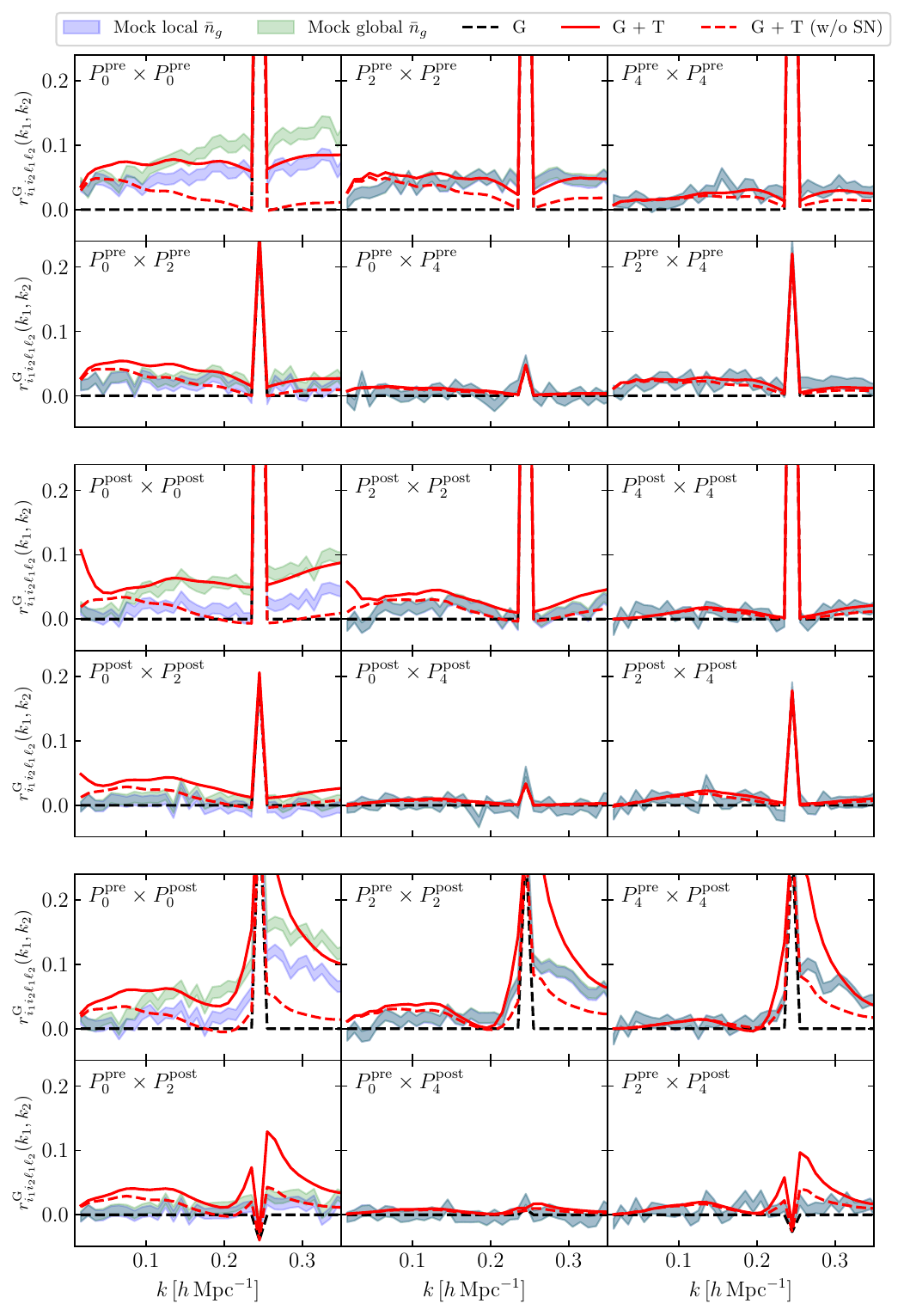}
\caption{Similar to Fig.~\ref{fig:baseline_rcovG_0.045}, but with $k_2=0.245\,h\,\mathrm{Mpc}^{-1}$.}
\label{fig:baseline_rcovG_0.245}
\end{figure}

To compare non-diagonal elements, we use symbol $r^\mathrm{G}_{i_1 i_2 \ell_1 \ell_2}(k_1, k_2)$ to represent the covariance matrix normalized by the Gaussian prediction
\begin{equation}
r^\mathrm{G}_{i_1 i_2 \ell_1 \ell_2}(k_1, k_2) \equiv \frac{C_{i_1 i_2 \ell_1 \ell_2}(k_1, k_2)}{\sqrt{C^\mathrm{G}_{i_1 i_1 \ell_1 \ell_1}(k_1, k_1) C^\mathrm{G}_{i_2 i_2 \ell_2 \ell_2}(k_2, k_2)}}\text{.}
\end{equation}
Figure~\ref{fig:baseline_rcovG_0.045}, \ref{fig:baseline_rcovG_0.095}, \ref{fig:baseline_rcovG_0.145}, \ref{fig:baseline_rcovG_0.195} and \ref{fig:baseline_rcovG_0.245} show $r^\mathrm{G}_{i_1 i_2 \ell_1 \ell_2}(k_1, k_2)$ at $k_2=0.045, 0.095, 0.145, 0.195, 0.245\,h\,\mathrm{Mpc}^{-1}$, respectively. In each plot, we also show the non-Gaussian analytic covariance with the shot noise contribution set to zero. Overall, the agreement between the theory prediction and the numerical covariance is reasonably well. Even when $k > 0.15\,h\mathrm{Mpc}^{-1}$, the theory prediction still captures the shape of correlations between different $k$-bins. The non-Gaussian effect in the cross-covariance is comparable to, or even stronger than the pre-reconstruction covariance. According to the theoretical calculation, we conclude that this enhancement is due to the shot noise effect.

We also notice there are a few features appearing in the analytic covariance but not present in the numerical covariance. For example, the $P_0^\mathrm{post}\times P_0^\mathrm{post}$ block in Fig.~\ref{fig:baseline_rcovG_0.245} shows peculiar enhancements in small $k$.
It is likely caused by the second line in eq(\ref{eq:B***}), which diverges in the squeezed limit when $\bs{k}_3$ goes to $\bs{0}$ because $P_{gg}^\NN(\bs{k}_3)\to\mathrm{const}$ while $R_2(\bs{k}_2, \bs{k}_3)\to\infty$. In addition, at small scales, the theory strongly over-predicts the correlation when $k_1$ approaches $k_2$ in the cross-correlation covariance. This may be attributed to the IR-effect as pointed by \citet{Sugiyama:2024eye}, i.e., the cross-power spectra and it's shot noise should damp exponentially due to the lack of IR-cancellation. However, deriving the full IR-resummation equation for the cross correlation trispectrum is beyond the scope of this paper, and we leave it to future work.

\begin{figure}
\centering
\includegraphics[width=1.0\linewidth]{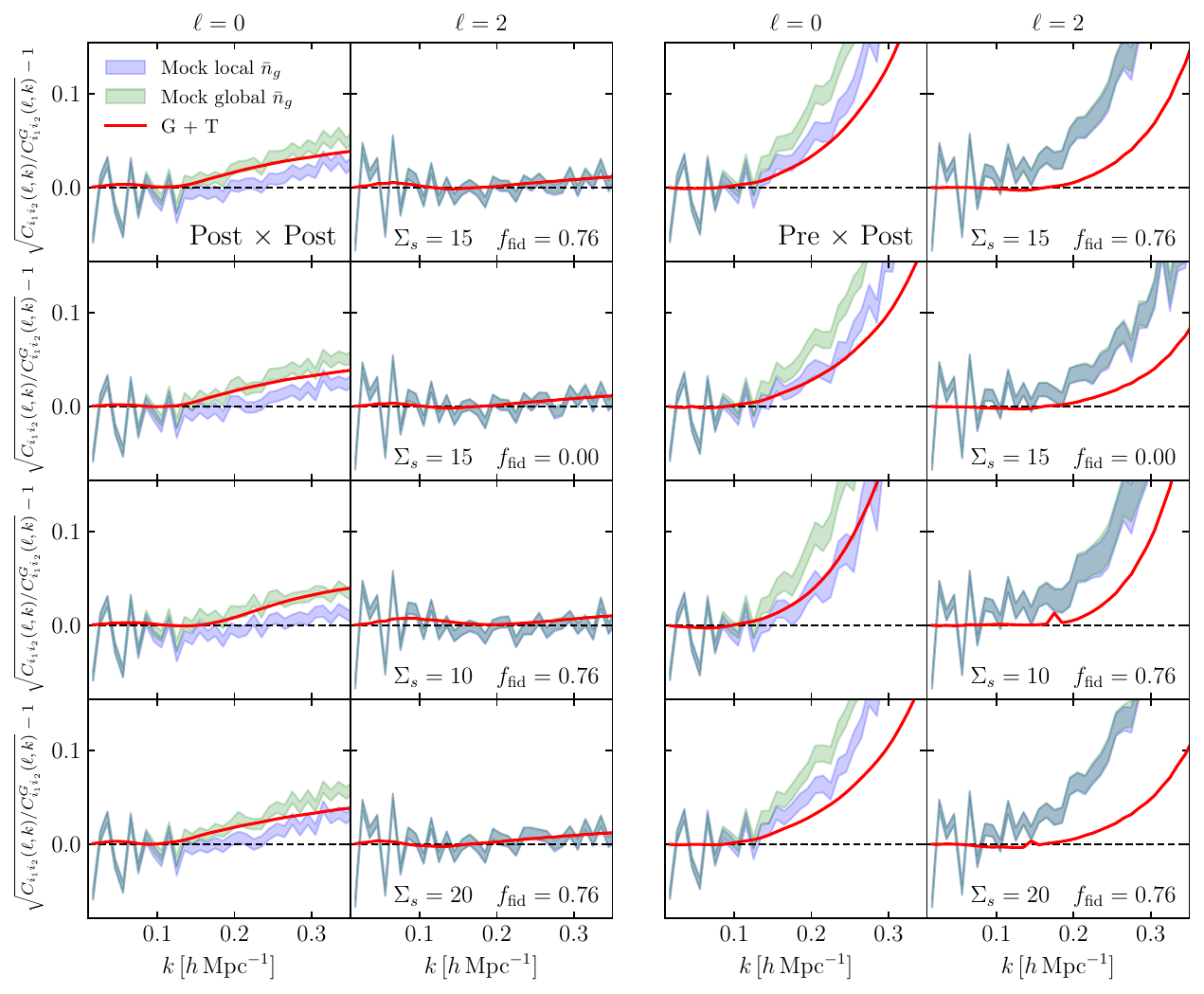}
\caption{Similar to Fig.~\ref{fig:baseline_diag}, but comparing different reconstruction settings.}
\label{fig:reconstruction-variants-diagonal-cov}
\end{figure}

\begin{figure}
\centering
\includegraphics[width=1.0\linewidth]{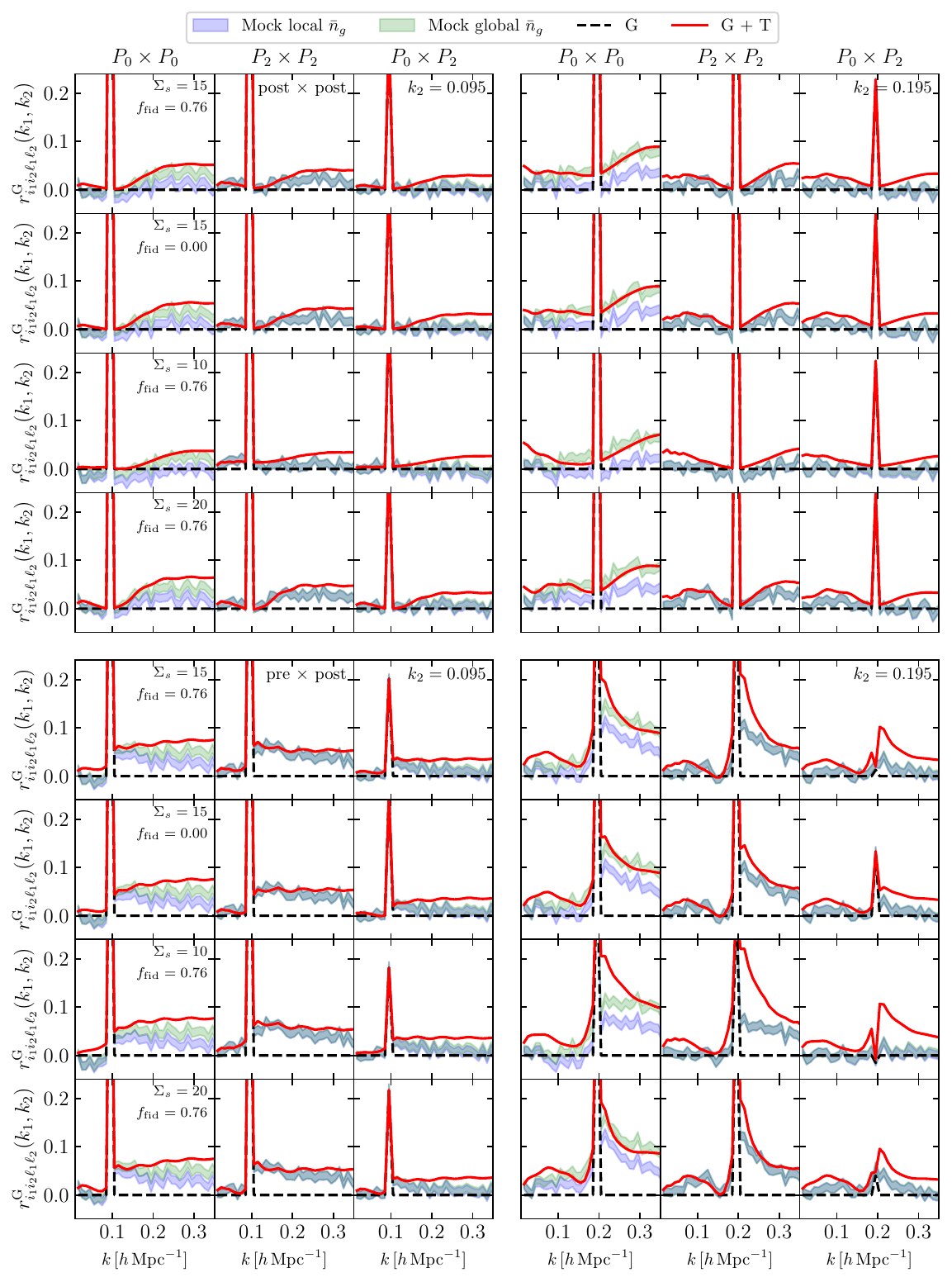}
\caption{Similar to Fig.~\ref{fig:baseline_rcovG_0.045}, but comparing different reconstruction settings.}
\label{fig:reconstruction-variants-nondiagonal-cov}
\end{figure}

We then study the change of numeric and analytic covariances by varying the smoothing scale and the fiducial growth rate. Results are shown in Fig.~\ref{fig:reconstruction-variants-diagonal-cov} and Fig.~\ref{fig:reconstruction-variants-nondiagonal-cov}. Changing $f_\mathrm{fid}$ mainly changes the quadruple contribution, and increasing the smoothing scale increases the correlations in the post-reconstruction power spectra while decreasing the correlations in the pre- and post-reconstruction cross covariance.  This is because the post-reconstruction power spectrum recovers the pre-reconstruction power spectrum when $\Sigma_s\to \infty$, so that $C_{gg**}$ reduces to $C_{gggg}$. The theory prediction is still in good agreement with the numeric covariance, and the theory matches the numerical covariance better when using a larger smoothing scale or a zero $f_\mathrm{fid}$.

\section{Conclusion and discussions}
\label{sec:con}

The reconstruction technique provides a useful tool to improve the measurement of the BAO signal and makes the standard ruler more robust by undoing the nonlinear gravitational evolution. This field level operation brings back information beyond the $2$-point statistics but also complicates the analysis especially when combining with the pre-reconstruction statistics. This is because both the signal and the covariance are affected by higher-order statistics.

In this work, we build a theory model of the covariance matrix for the power spectra before and after the BAO reconstruction based on the recent theoretical progress in the reconstruction modelling \citep{Sugiyama:2024eye, Sugiyama:2024ggt, Sugiyama:2024qsw}. We check the accuracy of our model against $15000$ halo mocks at redshift $z=0.5$. We show that the diagonal part of the covariance matrix is well described by the Gaussian prediction, with the maximum deviation of $5$ percent for the pre- and post-reconstruction auto covariance. However, the cross covariance between pre- and post-reconstruction power spectra deviates from the Gaussian prediction quickly when $k>0.1\,h\,\mathrm{Mpc}^{-1}$ and achieves $15$ per cent at $k=0.25\,h\,\mathrm{Mpc}^{-1}$. In addition, the non-Gaussian effect in the cross covariance is comparable to, or even stronger than the pre-reconstruction covariance. We find that adding the non-Gaussian covariance predicted by the perturbation theory leads to a better agreement with the numeric covariance up to $k\simeq 0.15\,h\,\mathrm{Mpc}^{-1}$, and the shot noise effect has a very large contribution at small scales. When $k > 0.15\,h\,\mathrm{Mpc}^{-1}$, the theory prediction still captures the shape of correlations between different $k$-bins, but over-predicts the correlation near the diagonal elements in the cross covariance. We also test the model with different smoothing scales and different fiducial growth rates when doing the reconstruction, and we find similar agreements with the numeric covariance.

In future work, we plan to extend the model to account for the IR-resummation effect. It will also be interesting to extend the calculation to multiple galaxy populations \citep{Mergulhao:2023zso, Zhao:2023ebp, Ebina:2024ojt}, correlation functions \citep{Philcox:2019ued, Rashkovetskyi:2023nwt} and compare the covariance matrix performance at the parameter level \citep{DESI:2024sbq, KP4s8-Alves, KP4s6-Forero-Sanchez}. The impact of the survey window function \citep{Wadekar:2019rdu}, the local average effect \citep{dePutter:2011ah}, the super survey effect \citep{Takada:2013wfa} and the fiber collision affects \citep{Pinon:2024wzd} on covariance matrices are also interesting topics.  This work provides a useful reference for future extensions.

\appendix
\section{Perturbation theory before reconstruction}
\label{appendix:perturbation}
The galaxy density field in redshift space can be expanded perturbatively
\begin{equation}
    \delta_g^{(n)}(\bs{k}) = \int_{\bs{k}=\bs{p}_{1\dots n}} Z_n(\bs{p}_1, \cdots, \bs{p}_n)\delta_L(\bs{p}_1)\cdots\delta_L(\bs{p}_n)\text{,}
\end{equation}
where we have adopted the EdS-approximation, with $\delta_L$ the linear matter density field evaluated at the observation redshift and $Z_n$ the $n$-th order perturbation kernel. In this work, we use the ``descendants'' bias basis defined in \citet{Perko:2016puo} and assumes $\tilde{c}_{\delta, 1} = b_1, \tilde{c}_{\delta, 2} = 1, \tilde{c}_{\delta, 3} = 1$, with all other bias parameters set to $0$. The tree-level bispectrum and trispectrum are given by \citep{Goroff:1986ep}
\begin{align}
B_{ggg}(\bs{k}_{1},\bs{k}_{2},\bs{k}_{3})&=2Z_{1}(\bs{k}_{1})Z_{1}(\bs{k}_{2})Z_{2}(\bs{k}_{1},\bs{k}_{2})P_{L}(k_{1})P_{L}(k_{2})+(\text{2 cyc.})\\
T_{gggg}(\bs{k}_{1},\bs{k}_{2},\bs{k}_{3},\bs{k}_{4})&=4Z_{1}(\bs{k}_{1})Z_{1}(\bs{k}_{2})Z_{2}(-\bs{k}_{1},\bs{k}_{14})Z_{2}(-\bs{k}_{2},\bs{k}_{23})P_{L}(k_{1})P_{L}(k_{2})P_{L}(k_{14})+(\text{11 perm.})\nonumber\\ &+6Z_{1}(\bs{k}_{1})Z_{1}(\bs{k}_{2})Z_{1}(\bs{k}_{3})Z_{3}(\bs{k}_{1},\bs{k}_{2},\bs{k}_{3})P_{L}(k_{1})P_{L}(k_{2})P_{L}(k_{3})+(\text{3 cyc.})\text{.}
\end{align}
The full power spectrum, bispectrum and trispectrum with the shot noise effect are given by \citet{Sugiyama:2019ike} and \citet{Sugiyama:2024qsw}. We copied these expressions in the following
\begin{align}
P_{gg}^{\NN}(\bs{k})&=P_{gg}(\bs{k})+\frac{1}{\bar{n}_{g}}\\
B_{ggg}^{\NN}(\bs{k}_{1},\bs{k}_{2},\bs{k}_{3})&=B_{ggg}(\bs{k}_{1},\bs{k}_{2},\bs{k}_{3})+\frac{1}{\bar{n}_{g}}\left[P_{gg}(\bs{k}_{1})+P_{gg}(\bs{k}_{2})+P_{gg}(\bs{k}_{3})\right]+\frac{1}{\bar{n}_{g}^{2}}\\
B_{ggg}^{\NN12}(\bs{k}_{1},\bs{k}_{2},\bs{k}_{3})&=B_{ggg}(\bs{k}_{1},\bs{k}_{2},\bs{k}_{3})+\frac{1}{\bar{n}_{g}}[P_{gg}(\bs{k}_{1})+P_{gg}(\bs{k}_{2})]\\
T_{gggg}^{\NN}(\bs{k}_{1},\bs{k}_{2},\bs{k}_{3},\bs{k}_{4})&=T_{gggg}(\bs{k}_{1},\bs{k}_{2},\bs{k}_{3},\bs{k}_{4})\nonumber\\
&+\frac{1}{\bar{n}_{g}}\left[B_{ggg}(-\bs{k}_{12},\bs{k}_{1},\bs{k}_{2})+B_{ggg}(-\bs{k}_{13},\bs{k}_{1},\bs{k}_{3})+B_{ggg}(-\bs{k}_{14},\bs{k}_{1},\bs{k}_{4})\right.\nonumber\\
&\left.\phantom{ggg}+B_{ggg}(-\bs{k}_{23},\bs{k}_{2},\bs{k}_{3})+B_{ggg}(-\bs{k}_{24},\bs{k}_{2},\bs{k}_{4})+B_{ggg}(-\bs{k}_{34},\bs{k}_{3},\bs{k}_{4})\right]\nonumber\\
&+\frac{1}{\bar{n}_{g}^{2}}\left[P_{gg}(\bs{k}_{1})+P_{gg}(\bs{k}_{2})+P_{gg}(\bs{k}_{3})+P_{gg}(\bs{k}_{4})+P_{gg}(\bs{k}_{12})+P_{gg}(\bs{k}_{13})+P_{gg}(\bs{k}_{14})\right]\nonumber\\
&+\frac{1}{\bar{n}_{g}^{3}}\text{.}
\end{align}

\normalem
\begin{acknowledgements}
RZ thanks Nathan Findlay for the inspiration of using $n_*$ to represent the post-reconstruction field. RZ, YW and GBZ are supported by National Key R\&D Program of China No. 2023YFA1607803, NSFC grants 11925303, and by the CAS Project for Young Scientists in Basic Research (No. YSBR-092). RZ is also supported by the Chinese Scholarship Council (CSC) and the University of Portsmouth. KK is supported by the STFC grant ST/W001225/1. YW is also supported by NSFC Grants (12273048, 12422301), by National Key R\&D Program of China No. 2022YFF0503404, by the Youth Innovation Promotion Association CAS, and by the Nebula Talents Program of NAOC. GBZ is also supported by science research grants from the China Manned Space Project with No. CMS-CSST-2021-B01, and the New Cornerstone Science Foundation through the XPLORER prize. Numerical computations were done on the Sciama High Performance Compute (HPC) cluster which is supported by the ICG, SEPNet and the University of Portsmouth. For the purpose of open access, the authors have applied a Creative Commons Attribution (CC BY) licence to any Author Accepted Manuscript version arising.
\end{acknowledgements}
  
\bibliographystyle{raa}
\bibliography{bibtex}

\end{document}